\documentclass[manuscript]{aastex63}
\usepackage[utf8]{inputenc}
\usepackage[T1]{fontenc}
\usepackage{amsmath,amssymb,placeins}
\usepackage[version=4]{mhchem}
\usepackage{comment}
\usepackage{appendix}

\usepackage{todonotes}

\graphicspath{{./}{figures/}}

\received{September 20, 2019}
\revised{October 29, 2019}
\accepted{October 30, 2019}
\submitjournal{ApJ}

\shorttitle{Efficient Production of S$_8$ in Interstellar Ices}
\shortauthors{Shingledecker et al.}
\begin{document}

\title{Efficient Production of S$_8$ in Interstellar Ices: The effects of cosmic ray-driven radiation chemistry and non-diffusive bulk reactions}

\correspondingauthor{Christopher N. Shingledecker}
\email{cns@mpe.mpg.de}

\author[0000-0002-5171-7568]{Christopher N. Shingledecker}
\affil{
Max-Planck-Institut für extraterrestrische Physik, \\
D-85748 Garching, Germany
}
\affiliation{Institute for Theoretical Chemistry \\ University of Stuttgart \\ Pfaffenwaldring 55, 70569, Germany}
\affiliation{Department of Chemistry 
University of Virginia 
Charlottesville, VA, USA}

\author{Thanja Lamberts}
\affiliation{
Leiden Institute of Chemistry, Gorlaeus Laboratories, Leiden University \\
P.O. Box 9502, 2300 RA Leiden, The Netherlands
}

\author{Jacob C. Laas}
\affiliation{
Max-Planck-Institut für extraterrestrische Physik, \\
D-85748 Garching, Germany
}

\author{Anton Vasyunin}
\affil{Ural Federal University,  Ekaterinburg, Russia}
\affil{Visiting Leading Researcher, Engineering Research Institute 'Ventspils International Radio
Astronomy Centre' of Ventspils University of Applied Sciences, 
 In\v{z}enieru 101, Ventspils LV-3601, Latvia}
 
\author{Eric Herbst}
\affil{Department of Chemistry 
University of Virginia 
Charlottesville, VA, USA}
\affil{Department of Astronomy 
University of Virginia 
Charlottesville, VA, USA}

\author{Johannes K\"{a}stner}
\affiliation{Institute for Theoretical Chemistry \\ University of Stuttgart \\ Pfaffenwaldring 55, 70569, Germany}

\author{Paola Caselli}
\affiliation{
Max-Planck-Institut für extraterrestrische Physik, \\
D-85748 Garching, Germany
}

%% Note that the \and command from previous versions of AASTeX is now
%% depreciated in this version as it is no longer necessary. AASTeX 
%% automatically takes care of all commas and "and"s between authors names.

%% AASTeX 6.2 has the new \collaboration and \nocollaboration commands to
%% provide the collaboration status of a group of authors. These commands 
%% can be used either before or after the list of corresponding authors. The
%% argument for \collaboration is the collaboration identifier. Authors are
%% encouraged to surround collaboration identifiers with ()s. The 
%% \nocollaboration command takes no argument and exists to indicate that
%% the nearby authors are not part of surrounding collaborations.

%% Mark off the abstract in the ``abstract'' environment. 
\begin{abstract}

In this work, we reexamine sulfur chemistry occurring on and in the ice mantles of interstellar dust grains, and report the effects of two new modifications to standard astrochemical models; namely, (a) the incorporation of cosmic ray-driven radiation chemistry and (b) the assumption of fast, non-diffusive reactions for key radicals in the bulk. Results from our models of dense molecular clouds show that these changes can have a profound influence on the abundances of sulfur-bearing species in ice mantles, including a reduction in the abundance of solid-phase H$_2$S and HS, and a significant increase in the abundances of OCS, SO$_2$, as well as pure allotropes of sulfur, especially S$_8$. These pure-sulfur species - though nearly impossible to observe directly - have long been speculated to be potential sulfur reservoirs and our results represent possibly the most accurate estimates yet of their abundances in the dense ISM. Moreover, the results of these updated models are found to be in good agreement with available observational data. Finally, we examine the implications of our findings with regard to the as-yet-unknown sulfur reservoir thought to exist in dense interstellar environments. 

\end{abstract}

%% Keywords should appear after the \end{abstract} command. 
%% See the online documentation for the full list of available subject
%% keywords and the rules for their use.

%% From the front matter, we move on to the body of the paper.
%% Sections are demarcated by \section and \subsection, respectively.
%% Observe the use of the LaTeX \label
%% command after the \subsection to give a symbolic KEY to the
%% subsection for cross-referencing in a \ref command.
%% You can use LaTeX's \ref and \label commands to keep track of
%% cross-references to sections, equations, tables, and figures.
%% That way, if you change the order of any elements, LaTeX will
%% automatically renumber them.
%%
%% We recommend that authors also use the natbib \citep
%% and \citet commands to identify citations.  The citations are
%% tied to the reference list via symbolic KEYs. The KEY corresponds
%% to the KEY in the \bibitem in the reference list below. 

\section{Introduction}

The recent detection of several new sulfur-bearing molecules in the interstellar medium (ISM) \citep{agundez_detection_2018,cernicharo_discovery_2018}, as well as in comet 67P/Churyumov-Gerasimenko (67P/C-G) by the \textit{Rosetta} orbiter \citep{calmonte_sulphur-bearing_2016} has, in part, spurred renewed interest in the chemistry of this malodorous element \citep{hudson_infrared_2018, morgan_astrophysical_2018,dungee_high-resolution_2018,danilovich_sulphur-bearing_2018,drozdovskaya_alma-pils_2018,zakharenko_deuterated_2019,gorai_search_2017}.  A number of recent studies have been motivated by the long-standing mystery regarding sulfur abundances in dense regions; namely, that though the total observed abundance of S-containing species in diffuse environments is approximately the cosmic value, in dense cores, sulfur appears to be depleted by up to several orders of magnitude \citep{prasad_sulfur_1982,jenkins_unified_2009,anderson_new_2013}. One possible explanation for this apparently ``missing'' sulfur is that a great deal of it is incorporated into some as yet unknown molecule trapped in dust-grain ice mantles. 

Over the past few years, quite a few attempts have been made to shed light on what this species might be using astrochemical models \citep{le_gal_sulfur_2019,laas_modeling_2019,vidal_new_2018,vidal_reservoir_2017,semenov_chemistry_2018,vastel_sulphur_2018}. For instance, based on results from their simulations of the cold core TMC-1, \citet{vidal_reservoir_2017} suggested that sulfur might exist mostly as either solid HS or \ce{H2S}, or as neutral atomic sulfur in the gas, depending on the age of the source. However, despite such predictions - and the fact that \ce{H2S} was recently detected by the \textit{Rosetta} orbiter - thus far OCS remains the only sulfur-bearing species definitively observed in interstellar ices \citep{palumbo_solid_1997}, though tentative detections of \ce{SO2} have also been reported by \citet{boogert_infrared_1997} and \citet{zasowski_spitzer_2009}.  More recently, \citet{laas_modeling_2019} - using a new chemical network that included a particularly comprehensive set of sulfur-related reactions - concluded that S was bound-up in organo-sulfur molecules in dust-grain ice mantles. More generally, though, there are a number of critical weaknesses in how current astrochemical codes simulate ice chemistry related to, e.g. chemical desorption, microscopic ice structure, reactant orientation, cosmic ray-driven processes, and the dominant mechanism of bulk reactions. Though all of these topics are worthy of further study, we will focus here on the final two.

Cosmic rays and other forms of ionizing radiation, related to the first major shortcoming we address in this work, are ubiquitous in extraterrestrial environments. Unlike photons, cosmic rays are not quickly attenuated in dense molecular clouds \citep{ivlev_penetration_2018,silsbee_magnetic_2018,padovani_production_2018}, and in fact drive much of the chemistry of these regions. Two major examples include the formation of \ce{H3+} following the ionization of molecular hydrogen \citep{herbst_formation_1973}, and the production of internal UV photons due to electronic excitation of \ce{H2} \citep{prasad_uv_1983}. Yet, as noted, interactions between these energetic particles and dust-grain ice mantles have until now only been considered in a very limited, approximate way, in spite of a large body of previous laboratory work, which has proven that the irradiation of interstellar ice-analogues by energetic particles can drive a rich chemistry at even very low temperatures. Including  such processes is likely important for accurately modeling solid-phase sulfur chemistry in dust-grain ice mantles. For example, experiments by \citet{ferrante_formation_2008} have shown that OCS readily forms in S-containing ices bombarded by 0.8 MeV protons. Connections between this ion-driven ``radiation chemistry'' and interstellar sulfur were further strengthened by the detection of \ce{S2} in the coma of comet 67P/C-G by \citet{calmonte_sulphur-bearing_2016}. In that work, the authors concluded that this \ce{S2} likely formed in the pre-solar nebula via the radiolysis of species such as \ce{H2S}. In an attempt to improve how astrochemical models treat non-thermal processes, particularly those driven by cosmic rays,  we have recently developed methods for including such radiation chemistry in rate-equation-based codes \citep{shingledecker_general_2018}. Our preliminary findings are that the addition of these new mechanisms generally improves the agreement between models and observations \citep{shingledecker_cosmic-ray-driven_2018}; however, given the novelty of our approach, no modeling studies have yet been done which focus on the effects of cosmic ray-bombardment on the abundances of sulfur-bearing species in ice mantles.

Another major source of uncertainty in current simulations of grain chemistry concerns reactions within the bulk of dust-grain ice mantles. One significant source of uncertainty in such models involves whether or to what degree bulk diffusion is important, and if so, what the underlying mechanism behind this diffusion might be. Commonly, models today typically assume, for example, swapping \citep{oberg_quantification_2009,fayolle_laboratory_2011} or diffusion via interstitial sites \citep{lamberts_water_2013,chang_interstellar_2014,shingledecker_new_2017}, where the energetic barriers to bulk diffusion, $E_\mathrm{b}^\mathrm{bulk}$, are taken to be some fraction of the desorption energy, $E_\mathrm{D}$, and are highly uncertain. 
In \citet{shingledecker_simulating_2019}, we attempted to reduce this ambiguity by simulating well-constrained experiments, rather than the ISM. In our preliminary studies reported there, we found that the assumption that radicals within ices react via thermal diffusion leads to generally poor agreement between calculated and empirical results, due to the much slower chemistry in the simulations than what is shown to occur in the lab. Conversely, good agreement with experimental data was achieved by assuming that radicals in the ice react predominantly with their nearest neighbors, i.e. non-diffusively. These results are  in qualitative agreement with a recent study by \citet{ghesquiere_reactivity_2018}, who concluded that true bulk diffusion does not occur, rather, as temperatures increase, bulk species can be ``passively'' transported due to structural changes such as pore collapse or crystallization, or can ``actively'' diffuse along internal surfaces or cracks.

This work, therefore, is an attempt to build upon these recent investigations, and to examine what effect (a) cosmic ray-driven radiation chemistry, and (b) the fast, non-diffusive reaction of radicals in the bulk have on the abundances of S-bearing species in simulations of dense cores, and moreover, what impact such additions have on the major sulfur reservoirs predicted by our models. The rest of the paper is organized as follows: in \S \ref{sec:model} we provide details regarding the code, chemical network, and underlying theory; in \S \ref{sec:results} we present our main results; and finally, the main conclusions of this study are summarized in \S \ref{sec:conc}. 

\section{Model and Theory} \label{sec:model}

\subsection{Astrochemical Code and Chemical Network}

In this work, we have utilized the rate equation-based astrochemical model, \texttt{MONACO} \citep{vasyunin_formation_2017}, which we have previously modified, as described in  \citet{shingledecker_simulating_2019}. Specifically, these modifications allow for (1) the inclusion of cosmic ray-driven radiation processes, including the formation and barrierless non-diffusive reaction of short-lived \textit{suprathermal} species using the method of  \citet{shingledecker_general_2018} and \citet{shingledecker_cosmic-ray-driven_2018}, and (2)  the non-diffusive reaction of \textit{thermal} radicals in the bulk of the ice. These modifications have been tested and shown to yield excellent agreement with experiments involving irradiated \ce{O2} and \ce{H2O} ices  \citep{shingledecker_simulating_2019}. 

In brief, the basis of the method described in \citet{shingledecker_general_2018} for irradiation begins with the assumption that, for any grain species, $A$, collision by an energetic particle can lead to one of the following outcomes:

\begin{equation}
  aA \leadsto aA^+ + e 
  \label{p1}
  \tag{P1}
\end{equation}

\begin{equation}
  aA \leadsto aA^+ + e \rightarrow aA^* \rightarrow bB^* + cC^*
  \label{p2}
  \tag{P2}
\end{equation}

\begin{equation}
  aA \leadsto aA^* \rightarrow bB + cC
  \label{p3}
  \tag{P3}
\end{equation}

\begin{equation}
  aA \leadsto aA^*,
  \label{p4}
  \tag{P4}
\end{equation}

\noindent
where $B$ and $C$ are products, and $a$, $b$, and $c$ are the stoichiometric coefficients. One can then calculate rate coefficients, $k$, for processes \eqref{p1}-\eqref{p4} using

\begin{equation}
	k_\mathrm{P1} = G_\mathrm{P1} \left(\frac{S_\mathrm{e}}{100\;\mathrm{eV}}\right) \left(\Phi_\mathrm{ST}\left[\frac{\zeta}{10^{-17}}\right]\right)
   \label{k1}
\end{equation}

\begin{equation}
	k_\mathrm{P2} = G_\mathrm{P2} \left(\frac{S_\mathrm{e}}{100\;\mathrm{eV}}\right) \left(\Phi_\mathrm{ST}\left[\frac{\zeta}{10^{-17}}\right]\right)
   \label{k2}
\end{equation}

\begin{equation}
	k_\mathrm{P3} = G_\mathrm{P3} \left(\frac{S_\mathrm{e}}{100\;\mathrm{eV}}\right) \left(\Phi_\mathrm{ST}\left[\frac{\zeta}{10^{-17}}\right]\right)
   \label{k3}
\end{equation}

\begin{equation}
k_\mathrm{P4} = G_\mathrm{P4} \left(\frac{S_\mathrm{e}}{100\;\mathrm{eV}}\right) \left(\Phi_\mathrm{ST}\left[\frac{\zeta}{10^{-17}}\right]\right).
   \label{k4}
\end{equation}

\noindent
where here, $\Phi_\mathrm{ST}$ is the integrated Spitzer-Tomasko cosmic ray flux (8.6 particles cm$^{-2}$ s$^{-1}$) \citep{spitzer_heating_1968}, $\zeta$ is the H$_2$ ionization rate, and $S_\mathrm{e}$ is the so-called electronic stopping cross section, for which we use a average value of  $S_\mathrm{e} = 1.287 \times 10^{-15}$ cm$^{-2}$ eV \citep{shingledecker_general_2018,shingledecker_cosmic-ray-driven_2018}. The suprathermal species produced via processes \eqref{p2} and \eqref{p4} are critical for accurately reproducing energetic particle-driven chemistry at low temperatures \citep{abplanalp_study_2016}. In our model, we assume that, once formed, they quickly ($\sim10^{-14}$ s) either react barrierlessly with a neighboring species or are quenched by the ice \citep{roessler_suprathermal_1991,shingledecker_cosmic-ray-driven_2018}. These rapid reactions involving suprathermal species will be considered as part of the radiolysis rather than post-radiolysis kinetics in the remainder of the paper, and are to be differentiated from rapid thermal reactions involving radicals in the bulk of the ice mantle.

Following \citet{vasyunin_formation_2017}, we assume the surface comprises the top four monolayers, from which species can desorb thermally, or via the non-thermal processes of photodesorption or chemical desorption - with the latter calculated using the formalism described in \citet{garrod_non-thermal_2007} and occurring with the standard desorption efficiency of $1\%$. \citet{oba_infrared_2018} found the chemical desorption of the reaction system \ce{H + H2S -> H2 + HS} and \ce{H + HS -> H2S} to have absolute lower and upper limits of 0.5\% and 60\%, respectively. The 1\% efficiency we employ here for the reaction \ce{H + HS -> H2S} in particular is thus on the lower end of the spectrum; however, a detailed investigation of chemical desorption in this context falls beyond the scope of this paper. The third phase of our model, the bulk, consists of all monolayers below the top four (the surface). In addition to the previously mentioned radiolysis processes we have added, described above, it is assumed in the base \texttt{MONACO} code that photodissociation and reactions between bulk species can occur \citep{vasyunin_formation_2017}.

Regarding tunneling through diffusion barriers, \citet{senevirathne_hydrogen_2017} and \citet{asgeirsson_efficient_2018} showed theoretically and Kuwahata et al. (2015) showed experimentally that, for atomic H, this effect is only important for crystalline water surfaces or at very low temperatures (< 10 K). However, results from \citet{asgeirsson_efficient_2018} suggest that there is a large spread in the rate coefficients, determined by the large range of binding energies on ASW surfaces, and thus, the most diffusive H atoms will be those bound lightly. To take the large spread in binding energies and resulting  diffusive rate constants into account, for instance via a bimodal energy distribution \citep{cuppen_modelling_2011}, falls beyond the scope of the current paper. Therefore, based on the rate coefficients calculated in both theoretical studies, we use a modified first-order rate coefficient for H hopping on the high end of the spectrum, i.e., $\sim4\times10^9$ s$^{-1}$ at 10 K. For tunneling through chemical activation barriers, we use the standard formalism of \citet{hasegawa_models_1992} for H and \ce{H2}.

Our chemical network is based on the one recently developed by \citet{laas_modeling_2019}. To this we have added a number of reactions, described below. For a full list of gas and grain reactions, see Table \ref{tab:network} in Appendix A.  The most substantial additions to the network consisted of cosmic-ray-driven processes, including the radiolytic destruction of grain species, and reactions involving the resulting products. For the radiolysis processes, we have added both those given in \citet{shingledecker_cosmic-ray-driven_2018} and \citet{shingledecker_simulating_2019}, as well as new radiolytic destruction routes for sulfur-bearing species - including pure-sulfur species from \ce{S2} to \ce{S8} - with rate coefficients estimated using the method of \citet{shingledecker_general_2018}. We have also included a number of solid-phase reactions identified in previous experimental studies of irradiated sulfur-bearing ices \citep{jimenez-escobar_sulfur_2011,ferrante_formation_2008,moore_radiolysis_2007,chen_formation_2015}.

\subsubsection{Revised Competition Formula for Surface Reactions}

\begin{deluxetable*}{lccccl}[bt!]
\tablecaption{Reactions updated based on results from ab initio calculations by \citet{lamberts_tunneling_2017} (LK17) and \citet{lamberts_interstellar_2018} (L18). \label{tab:tunnel}}
\tablehead{
\colhead{Reaction} &
\colhead{$\alpha$} &
\colhead{$\beta$} & 
\colhead{$\gamma$} & 
\colhead{$T_0$} &
\colhead{Source} \\ 
\colhead{} &
\colhead{(s$^{-1}$)} &
\colhead{} & 
\colhead{(K)} & 
\colhead{(K)} & 
\colhead{}
}
\startdata
\ce{H + H2S -> H2 + HS} & $2.2\times10^{11}$ & 0.48 & 1400 & 180 & LK17 \\
\ce{H + CS -> HCS} & $3.3\times10^{11}$ & 0.50 & 100 & 35 & L18 \\
\ce{H + H2CS -> H2 + HCS} & $3.6\times10^{9}$ & 0.95 & 1295 & 145 & L18 \\
\ce{H + H2CS -> CH3S} & $1.6\times10^{11}$ & 0.50 & 290 & 85 & L18 \\
\ce{H + H2CS -> CH2SH} & $6.4\times10^{11}$ & 0.50 & 30 & 65 & L18 \\
\ce{H + CH3SH -> H2 + CH2SH} & $1.2\times10^{9}$ & 1.20 & 1710 & 155 & L18 \\
\ce{H + CH3SH -> H2 + CH3S} & $4.5\times10^{10}$ & 0.50 & 380 & 85 & L18 \\
\ce{H + CH3SH -> H2S + CH3} & $2.9\times10^{10}$ & 0.40 & 1060 & 70 & L18 \\
\enddata
\end{deluxetable*}

Given the low temperatures of many interstellar environments, quantum mechanical tunneling through reaction activation energies (barriers) is a particularly attractive mechanism when considering the abundances of interstellar molecules. Thus, we have also included and/or updated a number of surface reactions involving sulfur-bearing species, given in Table \ref{tab:tunnel}, which occur via tunneling at low temperatures. In so doing, we have also updated how \texttt{MONACO} handles tunneling and competition more generally, using the updated theory presented below. 

Currently, following \citet{hasegawa_models_1992}, it is common to multiply $R_{AB}$, the rate of diffusive surface reactions between some two species, $A$ and $B$, by a factor, $\kappa_{AB}$, which characterizes the probability of reaction. For exothermic barrierless reactions $\kappa_{AB}=1$, while for exothermic reactions with some activation energy, $E_\mathrm{a}$, $\kappa_{AB}\in[0,1]$.  The probability of overcoming $E_{a}$ (in units of K) per pass thermally is simply

\begin{equation}
\kappa^{AB}_\mathrm{therm} = \mathrm{exp}\left( 
-\frac{E_\mathrm{a}}{T},
\right)
\label{kappatherm}
\end{equation}

\noindent
where $T$ is the temperature. However, if there is at least one light reactant, e.g. H or H$_2$, or a process involving the transfer of H from one molecule to another, there is a chance that the reaction could proceed more efficiently at low temperatures via tunneling. Following \citet{tielens_model_1982}, one can approximate this probability by assuming a rectangular barrier of height $E_\mathrm{a}$ and width $a$: 

\begin{equation}
\kappa^{AB}_\mathrm{tunn} = \mathrm{exp}\left[ 
-2 \left( \frac{a}{\hbar} \right)
\sqrt{2\mu E_a}
\right].
\label{kappatunn}
\end{equation}

\noindent
Here, $\mu$ is a reduced or effective mass of the reactants and $a$ is commonly assumed to
be 1 \AA. A more realistic value of $\kappa_{AB}$ for tunneling was derived by
\citet{herbst_chemistry_2008}, which treats the competition between tunneling
and diffusion, i.e.

\begin{equation}
\kappa^{AB}_\mathrm{HM} = \frac{k_\mathrm{tunn}}{k_\mathrm{tunn} + k_\mathrm{hop,A} + k_\mathrm{hop,B}}
\label{kappahm}
\end{equation}

\noindent
where $k_\mathrm{hop,A}$ and $k_\mathrm{hop,B}$ are the hopping rates for $A$
and $B$, respectively \citep{herbst_chemistry_2008}, given by

\begin{equation}
    k_\mathrm{hop,A} = \nu_0^\mathrm{A} \mathrm{exp}\left(-\frac{E_\mathrm{b}^\mathrm{A}}{T} \right)
\end{equation}

and the first-order tunneling rate coefficient is given by 

\begin{equation}
k_\mathrm{tunn} = (\nu_0^\mathrm{A} + \nu_0^\mathrm{B}) \times \kappa^{AB}_\mathrm{tunn},
\label{ktunn}
\end{equation}

\noindent
with $\nu_0$ in the above expressions being the well-known attempt frequency
\citep{landau_mechanics_1976}, which typically has a value on the order of
$10^{12}$ s$^{-1}$ \citep{herbst_chemistry_2008} for physisorption.

\begin{figure}[t]
\centering
\hspace{-0.5cm}
\includegraphics[width=0.9\textwidth]{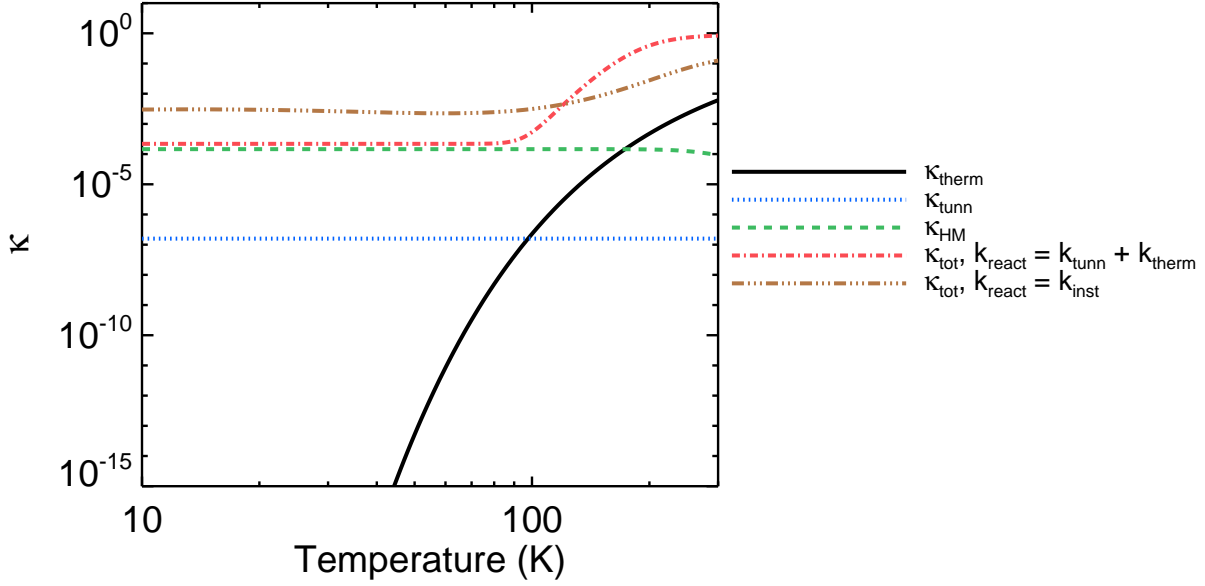}
\caption{Values of $\kappa$ calculated using a number of methods (see text) for the surface reaction $\mathrm{H} + \mathrm{H_2S} \rightarrow \mathrm{H_2} + \mathrm{HS}$. Here, the dot-dashed line uses $k_\mathrm{react}=k_\mathrm{tunn} + k_\mathrm{therm}$, whereas the triple dot-dashed line uses $k_\mathrm{react} = k_\mathrm{inst}$. Note: parameters for instanton results were taken from \citet{lamberts_tunneling_2017}.}
\label{fig:competition}
\end{figure}

In astrochemical models, a common practice is to compare $\kappa^{AB}_\mathrm{therm}$ with either $\kappa^{AB}_\mathrm{tunn}$ or $\kappa_\mathrm{HM}^{AB}$ at every temperature, and to select the largest of these \citep{garrod_three-phase_2013,vasyunin_unified_2013,taquet_multilayer_2012}.  Shown in Fig. \ref{fig:competition} are these values over 10-150 K for the surface reaction 

\begin{equation}
\mathrm{H} + \mathrm{H_2S} \rightarrow \mathrm{H_2} + \mathrm{HS}
\label{r1}
\end{equation}

\noindent
where, for illustration, we have utilized an activation energy of 1530 K \citep{lamberts_tunneling_2017} and assumed both reactants encounter one another via thermal diffusion.

Since astrochemical networks typically do not include separate reactions for the formation of products via the thermal or tunneling mechanism, however, the ultimate question of interest in models is not the mechanism by which a given reaction proceeds, but rather, whether it proceeds at all at a given temperature, and if so, with what rate. Thus, we here propose a new formalism for determining the probability of reaction, given simply by

\begin{equation}
  \label{kappanew}
  \kappa_\mathrm{tot}^{AB} = \frac{k_\mathrm{react}}{k_\mathrm{react} + k_\mathrm{hop,A} + k_\mathrm{hop,B}}
\end{equation}

\noindent

\begin{equation}
  k_\mathrm{react}=k_\mathrm{tunn} + k_\mathrm{therm}.    
  \label{kreact}
\end{equation}

\noindent
Thus, $\kappa_\mathrm{tot}^{AB}$ gives the total probability that a given reaction with a barrier will proceed - either thermally or via tunneling - when there is a competing chance that the reactants can diffuse away from each other. One advantage of Eq. \eqref{kappanew} is that it eliminates the need to explicitly compare the thermal vs. nonthermal probabilities at every temperature in the model, thereby reducing computational expense. Moreover, in simulations of hot core or shocks, use of Eq. \eqref{kappanew} will reduce potential discontinuities that can detrimentally effect numerical convergence. Since desorption is typically treated as a separate, distinct process in many astrochemical models, we do not include a desorption rate in the denominator of Eq. \eqref{kappanew} to prevent double counting of the phenomenon. We should note that one disadvantage of Eq. \eqref{kappatunn} is that, if $k_\mathrm{tunn} = k_\mathrm{therm}$, $k_\mathrm{react}$ will be a factor of 2 too large, however, the effects on the total result will likely be negligible.

For the tunneling factor in Eq. \eqref{kappanew}, one can use either the standard square potential, given in Eq. \eqref{kappatunn}, or alternatively, one can employ more realistic values, such as those obtained via instanton theory \citep{rommel_locating_2011,kastner_theory_2014}, which has recently been used to fruitfully study a number of reactions of astrochemical interest \citep{shingledecker_case_2019}. In one such study by \citet{lamberts_tunneling_2017}, the role of tunneling in reaction \eqref{r1} was investigated. For the reaction 

\begin{equation}
A + B \rightarrow [A \cdots B] \rightarrow C + D,
\end{equation}

\noindent
the first-order decay rates for the pre-reaction complex (PRC), $[A\cdots B]$, as described in \citet{lamberts_quantum_2016}, can be fit to a modified Arrhenius equation originally proposed by \citet{zheng_kinetics_2010} 

\begin{equation}
  k_\mathrm{inst} = \alpha \left(\frac{T}{300\;\mathrm{K}}\right)^{\beta} \mathrm{exp}\left(-\gamma\frac{T + T_0}{T^2 + T_0^2}\right) \mathrm{s^{-1}}
\label{kinstanton}
\end{equation}

\noindent
which can account for the approximately constant value of the rate coefficient at low temperatures due to tunneling. The resulting rate for the formation of products $C$ and $D$ from the PRC can be expressed as

\begin{equation}
    \frac{d[A\cdots B]}{dt} = -k_\mathrm{inst}[A \cdots B].
\end{equation}

\noindent
When available, one can use the results of such detailed calculations in Eq. \eqref{kappanew}, as shown in Fig. \ref{fig:competition} for reaction \ref{r1}. 

Here, it is important to stress two points regarding the kinds of reactions which proceed via the diffusive Langmuir-Hinshelwood mechanism, and for which the use of a competition formula is appropriate, namely that (1) the value of $\kappa$ cannot exceed unity, and (2) that regardless of the efficiency of overcoming the reaction barrier, the reactants still must encounter one another on the grain surface. Consequently, in the limit of highly efficient thermal activation over or tunneling through the reaction barrier, the rate of an exothermic reaction with a barrier approaches that of a barrierless exothermic reaction. 

Finally, for thermal bulk reactions which can proceed via tunneling, we have employed a different expression than Eq. \eqref{kappanew}, since it is assumed in that formula that the reactants have encountered one another diffusively, and there exists a nonzero chance that they can, given the presence of a finite chemical activation barrier, diffuse away from each other. For the bulk, we do not wish to assume that such diffusion occurs. Thus, we employ the following expression for $\kappa^\mathrm{AB}$ in cases where there is evidence of tunneling: 

\begin{equation}
    \kappa_\mathrm{bulk}^\mathrm{AB} = \frac{k_\mathrm{react}}{\nu_0^\mathrm{A}+\nu_0^\mathrm{B}}.
    \label{kappabulk}
\end{equation}

\noindent
Here, as with Eq. \eqref{kreact}, the numerator can be either the sum of a low-temperature and high-temperature term, or as in Eq. \eqref{kinstanton}, a single temperature-dependent expression that accounts for both. In Eq. \eqref{kappabulk}, $k_\mathrm{react}$ characterizes the rate at which the pre-reaction complex decays, and the denominator, the sums of the characteristic frequencies, approximates the number of times per second this complex can do so - with the quotient of the two thus accounting for the average probability that any such attempt will result in a successful reaction.

\subsection{Physical Conditions and Model Details}

\begin{deluxetable}{lc}[tb]
\tablecaption{Physical conditions \label{tab:physparameters}}
\tablewidth{0pt}
\tablehead{
\colhead{Parameter} & \colhead{Value}   
}
%\decimalcolnumbers
\startdata
$n_\mathrm{gas}$ & $10^4$ cm$^{-3}$ \\
$A_\mathrm{V}$ & 10 mag\\
$\zeta$ & $1.3\times10^{-17}$ s$^{-1}$ \\ 
$T_\mathrm{gas}$ & 10 K \\
$T_\mathrm{dust}$ & 10 K \\
\enddata
\end{deluxetable}

\begin{deluxetable}{lDc}[tb]
\tablecaption{Initial elemental abundances \label{tab:initialelemental}}
\tablewidth{0pt}
\tablehead{
\colhead{Element} & \twocolhead{Relative Abundance} & \colhead{Source} 
}
%\decimalcolnumbers
\startdata
\decimals
H & 0.9999 & \\
\ce{H2} & $5.0000\times10^{-5}$ & \\
\ce{He} & $9.5500\times10^{-2}$ & \citet{przybilla_cosmic_2008} \\
O       & $5.7544\times10^{-4}$ & \citet{przybilla_cosmic_2008} \\
\ce{C+} & $2.0893\times10^{-4}$ & \citet{przybilla_cosmic_2008} \\
N       & $5.7544\times10^{-5}$ & \citet{przybilla_cosmic_2008} \\
\ce{Mg+} & $3.6308\times10^{-5}$ & \citet{przybilla_cosmic_2008} \\
\ce{Si+} & $3.1623\times10^{-5}$ & \citet{przybilla_cosmic_2008} \\
\ce{Fe+} & $2.7542\times10^{-5}$ & \citet{przybilla_cosmic_2008} \\
\ce{S+} & $1.6600\times10^{-5}$ & \citet{esteban_reappraisal_2004} \\
\ce{Na+} & $1.7400\times10^{-6}$ & \citet{asplund_chemical_2009} \\
\ce{Cl+} & $2.8800\times10^{-7}$ & \citet{esteban_reappraisal_2004} \\
\ce{P+} & $2.5700\times10^{-7}$ & \citet{asplund_chemical_2009} \\
\ce{F} & $3.6300\times10^{-8}$ & \citet{asplund_chemical_2009} \\
\enddata
\end{deluxetable}

In order to investigate the role of radiation chemistry and non-diffusive bulk reactions on the abundances of sulfur species in dense interstellar regions, we have run simulations of generic cold cores using physical conditions and initial elemental abundances taken from \citet{laas_modeling_2019}, and listed in Tables \ref{tab:physparameters} and \ref{tab:initialelemental}.  

We have carried out five sets of simulations, as listed in Table 4.  Of these five, the results of two (A and B) are discussed here in detail and the results of the three others (C, D, and E) are to be found in Appendix B.   Our ``fiducial'' model, here referred to as Model A, uses standard treatments for surface and bulk chemistry as described in \citet{vasyunin_formation_2017}. Conversely, in Model B, we enable (\textit{i}) cosmic-ray-driven radiation chemistry as described in \citet{shingledecker_cosmic-ray-driven_2018}, including both the radiolytic destruction of grain mantle species and the subsequent production of electronically excited suprathermal species and thermal fragments, including radicals, (\textit{ii}) a non-diffusive bulk-reaction mechanism for a number of key bulk radicals relevant to our investigation and, finally, (\textit{iii}), the modified competition formula proposed here. Given the novelty of (\textit{ii}) in the context of astrochemical modeling, we here take an incremental approach to its implementation in our code as an initial test of its effect on the overall chemistry and composition of dust-grain ice mantles. In particular, we restrict the number of species that react via this non-diffusive mechanism to either those previously shown to be well-modeled in this way \citep{shingledecker_simulating_2019}, or most directly relevant for the grain chemistry of sulfur-bearing species, specifically, O, OH, \ce{HO2}, \ce{HO3}, H, HS, NS, HSO, C, S, and CS.

\begin{deluxetable}{lcccccc}[tb]
\tablecaption{Processes enabled in the simulations carried out for this work.  \label{tab:models}}
\tablewidth{0pt}
\tablehead{
\colhead{Process} & \colhead{Model A} & \colhead{Model B} & \colhead{Model C} & \colhead{Model D} & \colhead{Model E}     
}
%\decimalcolnumbers
\startdata
Cosmic-ray-driven Chemistry & \textit{no}  & \textit{yes} & \textit{no} & \textit{yes} & \textit{no} \\
Non-diffusive Bulk Reactions & \textit{no} & \textit{yes} & \textit{yes} & \textit{no} & \textit{no} \\
Modified Competition Formula & \textit{no} & \textit{yes} & \textit{yes} & \textit{yes} & \textit{yes}\\
\enddata
\end{deluxetable}

\FloatBarrier
\section{Results and Discussion} \label{sec:results}

\begin{deluxetable*}{cccccc}[bt!]
\tablecaption{Total calculated grain abundances (surface + bulk) of sulfur-bearing species relative to \ce{H2O} at $t\approx2\times10^6$ yr, as well as derived ice abundances for comet 67P/C-G taken from \citet{calmonte_sulphur-bearing_2016}. \label{tab:comet}}
\tablewidth{0pt}
\tablehead{
\colhead{Molecule} & \colhead{Model A} & \colhead{} & \colhead{Model B} & \colhead{} & \colhead{67P/Churyumov-Gerasimenko} 
%\colhead{}         & \colhead{}        & \colhead{}        &  \colhead{3.15 au, summer-winder}  & \colhead{$\sim$1 au} & \colhead{$\sim$1 au} & \colhead{Type} &
}
%\decimalcolnumbers
\startdata
\ce{H2S} & $3\times10^{-2}$  & & $4\times10^{-3}$ & & $(1.10 \pm 0.05)\times10^{-2}$\\
\ce{OCS} & $4\times10^{-5}$ & & $5\times10^{-4}$ & & $(4.08\pm0.09)\times10^{-4}$  \\
\ce{SO}  & $2\times10^{-7}$ & & $3\times10^{-8}$ & & $(7.1\pm1.1)\times10^{-4}$ \\
\ce{SO2}  & $4\times10^{-9}$ & &$7\times10^{-4}$ & & $(1.27\pm0.03)\times10^{-3}$  \\
\ce{CS2}  & $2\times10^{-9}$ & & $7\times10^{-8}$ & & $(5.68\pm0.18)\times10^{-5}$ \\
\ce{S2}   & $6\times10^{-12}$  & & $2\times10^{-7}$ & & $(1.97\pm0.35)\times10^{-5}$ \\
\ce{CH3SH}   & $3\times10^{-4}$ & & $1\times10^{-4}$ & & $(2.85\pm1.11)\times10^{-5}$\tablenotemark{a} \\
\ce{H2CS}   & $1\times10^{-6}$ & & $1\times10^{-6}$ & & $(2.67\pm0.75)\times10^{-6}$\tablenotemark{b} \\
\enddata
\tablenotetext{a}{2014 October - Table 4 of \citet{calmonte_sulphur-bearing_2016}}
\tablenotetext{b}{28.03.2015 12:14 - Table 5 of \citet{calmonte_sulphur-bearing_2016}}
%\tablecomments{Unless otherwise noted, cometary abundance values are taken from Table 3 of \citet{calmonte_sulphur-bearing_2016}.}
\end{deluxetable*}

\begin{figure*}
\gridline{
          \fig{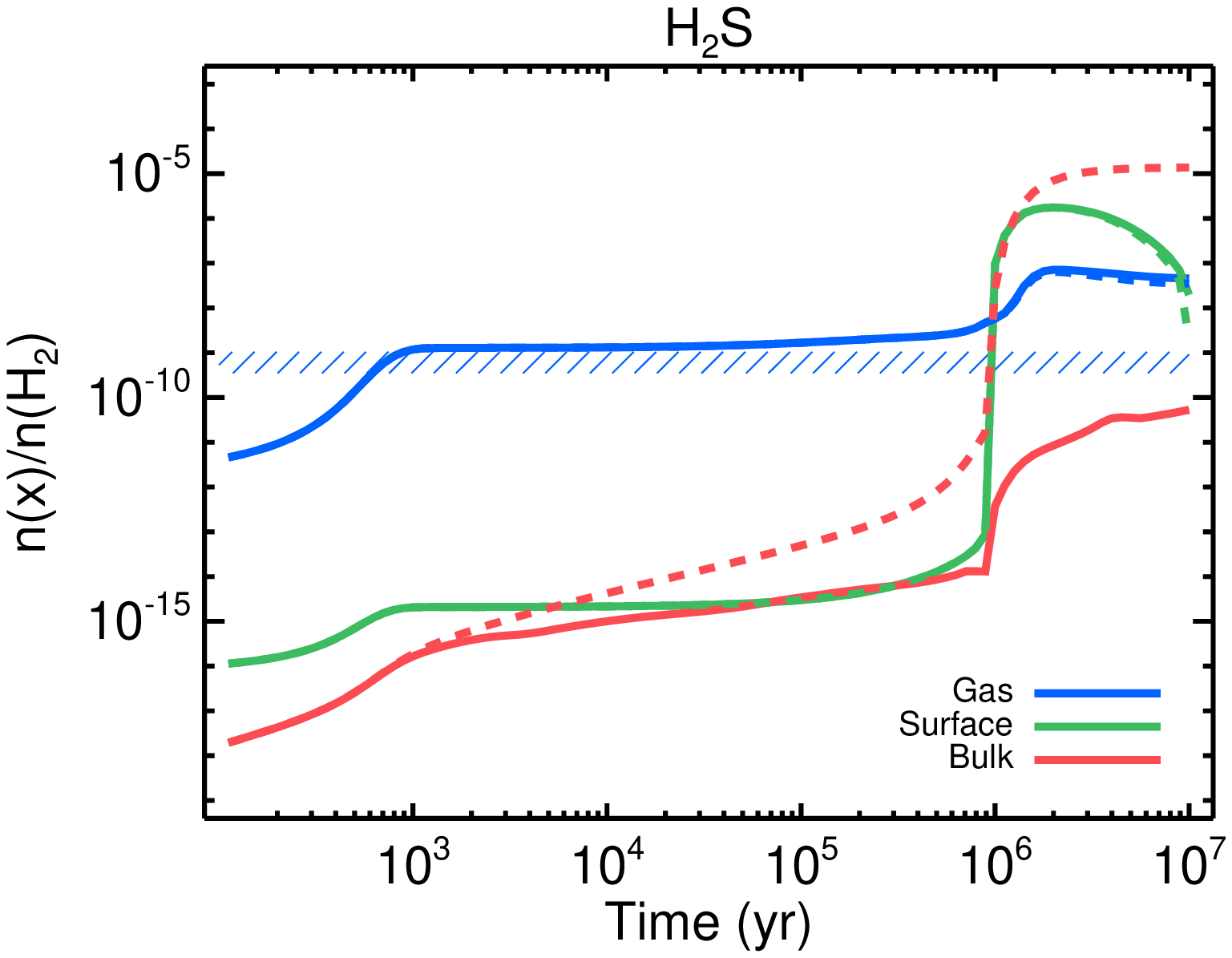}{0.5\textwidth}{}
          \fig{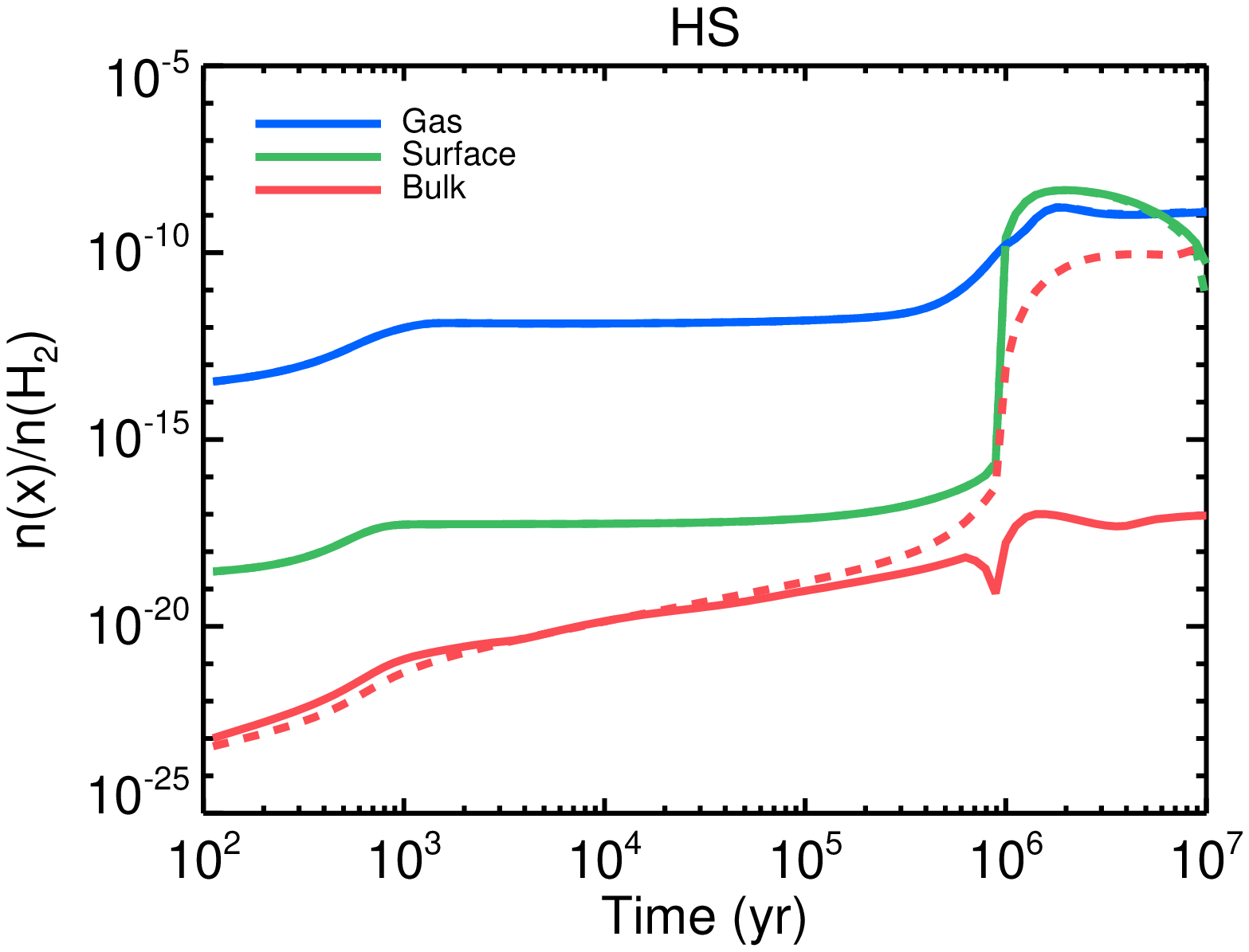}{0.5\textwidth}{}
          }
\gridline{
          \fig{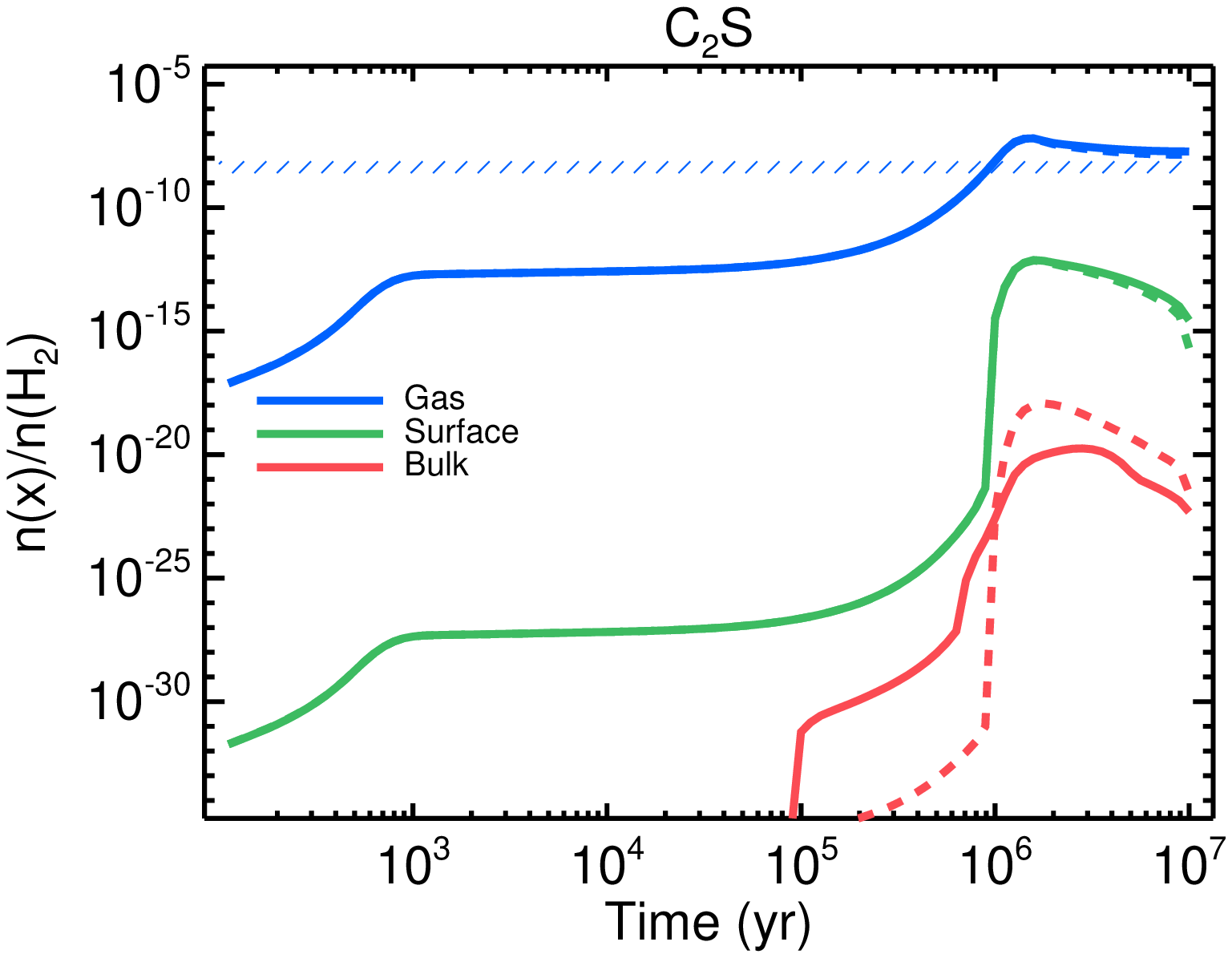}{0.5\textwidth}{}
          \fig{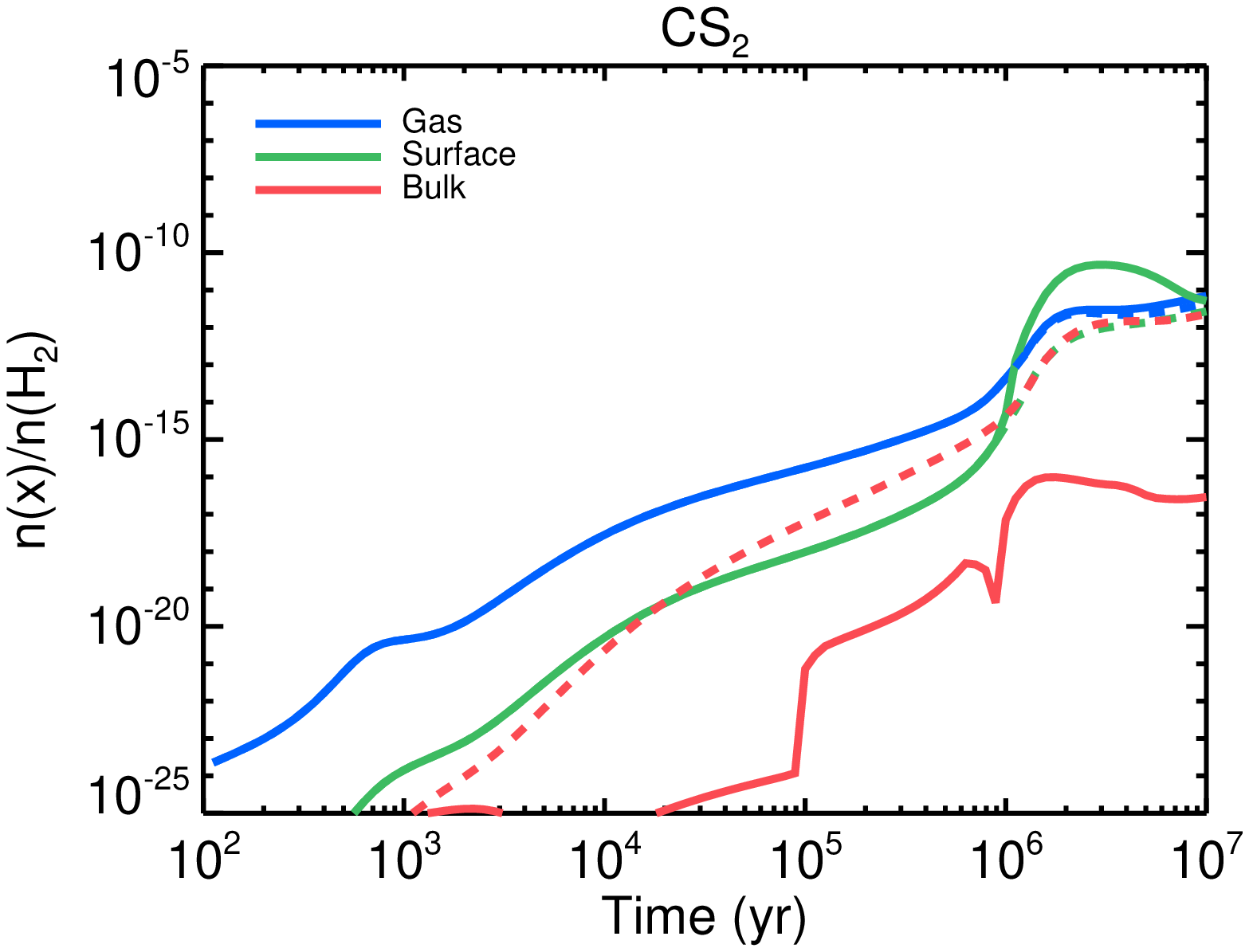}{0.5\textwidth}{}
}    
\gridline{
          \fig{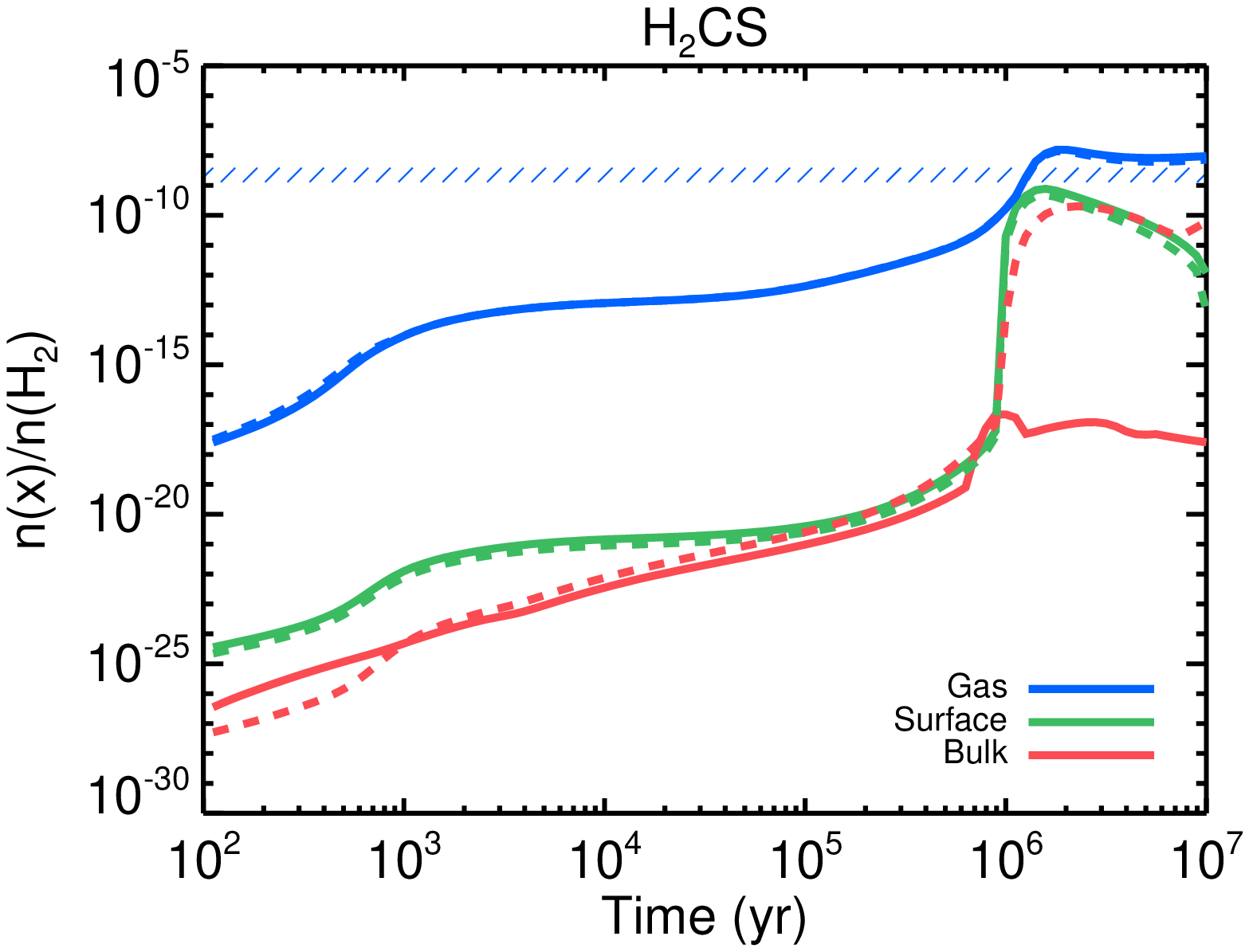}{0.5\textwidth}{}
          \fig{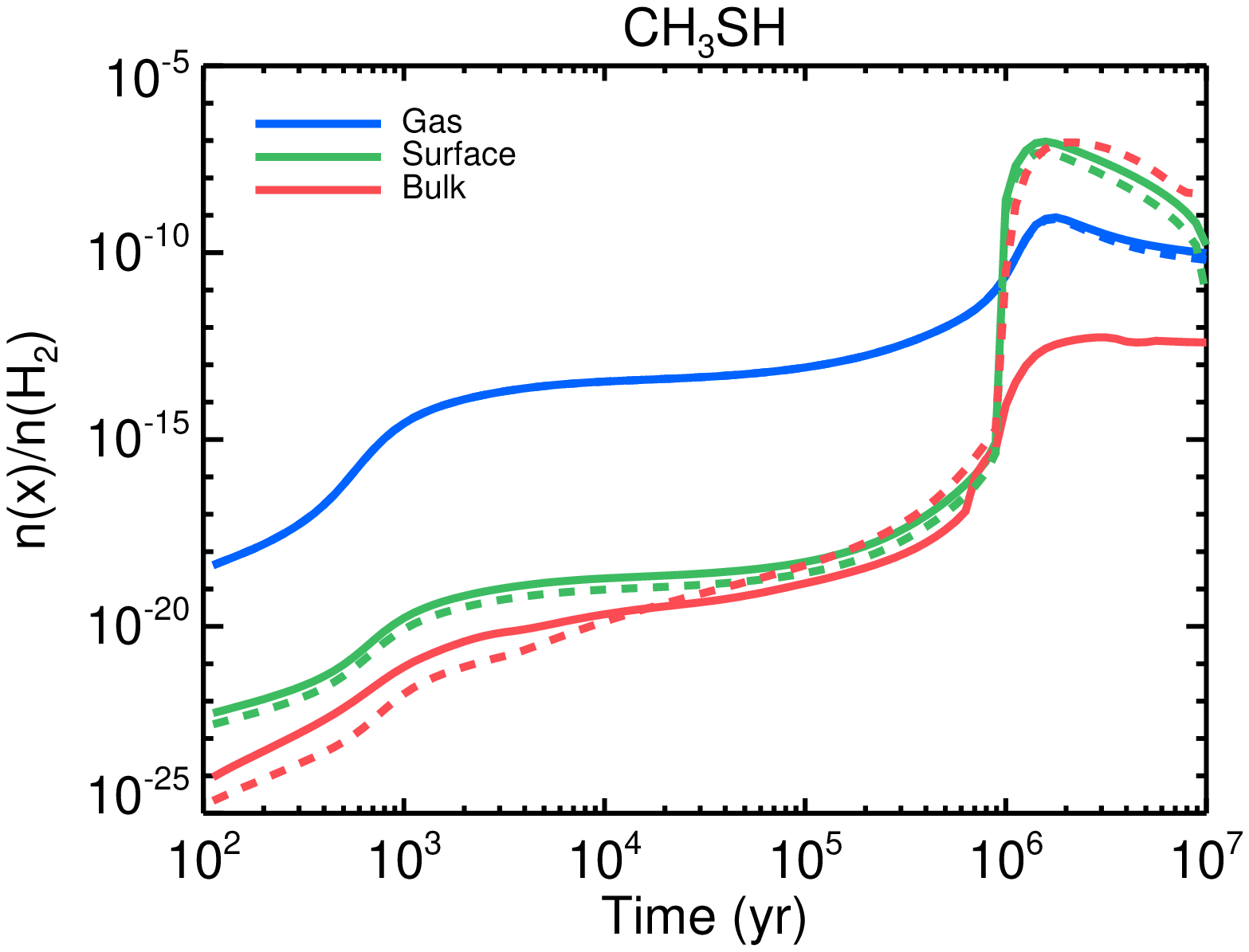}{0.5\textwidth}{}
}

\caption{ Calculated abundances of \ce{H2S}, \ce{HS}, \ce{H2CS}, \ce{C2S}, \ce{CS2}, and \ce{CH3SH} in models A (dashed line) and B (solid line). Gas-phase observational abundances for dense clouds, where available, are represented by horizontal blue hatched bars (see \citet{laas_modeling_2019} and references therein.).
\label{fig:block1}}
\end{figure*}

\begin{figure*}
\gridline{
          \fig{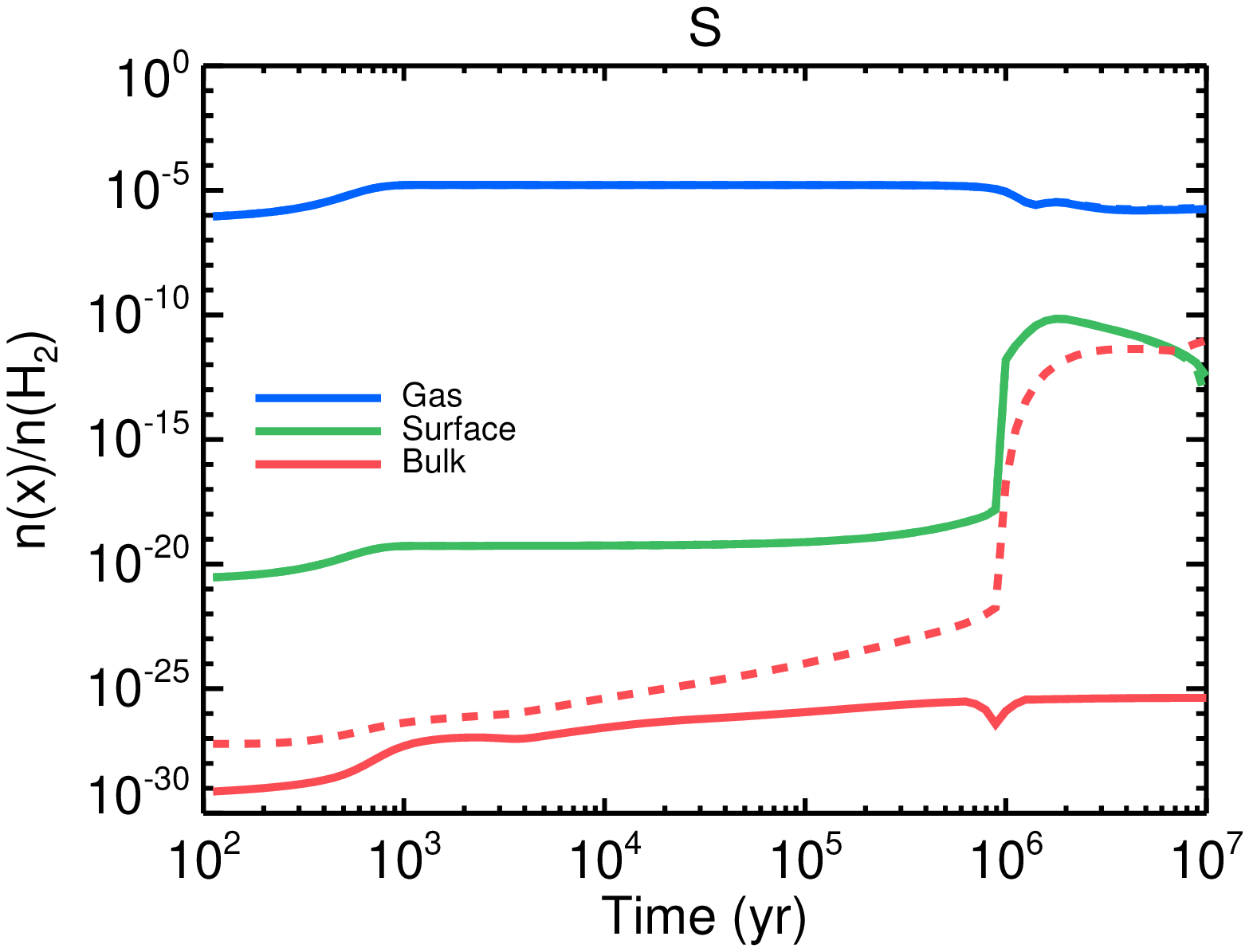}{0.5\textwidth}{}
          \fig{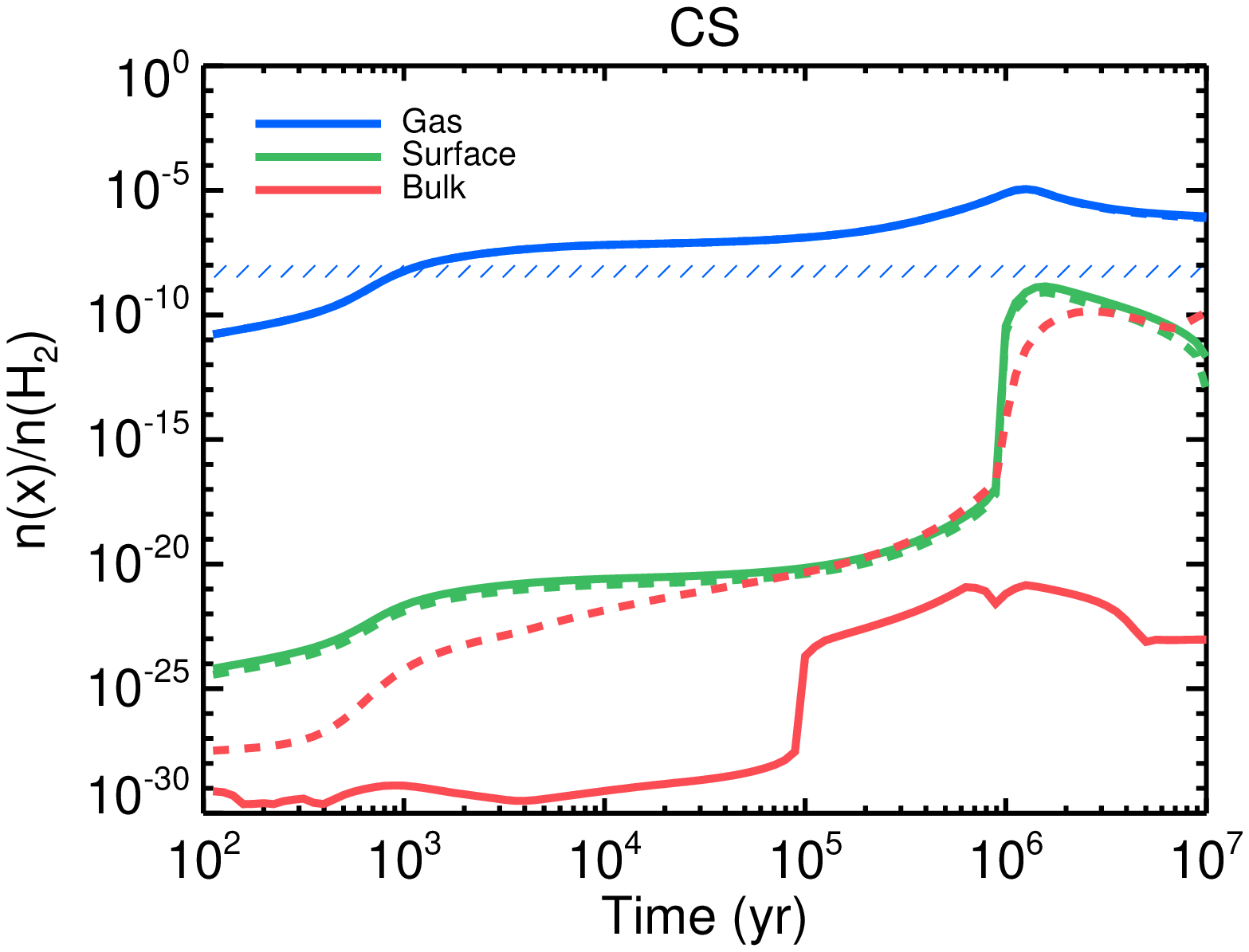}{0.5\textwidth}{}
}
\gridline{
          \fig{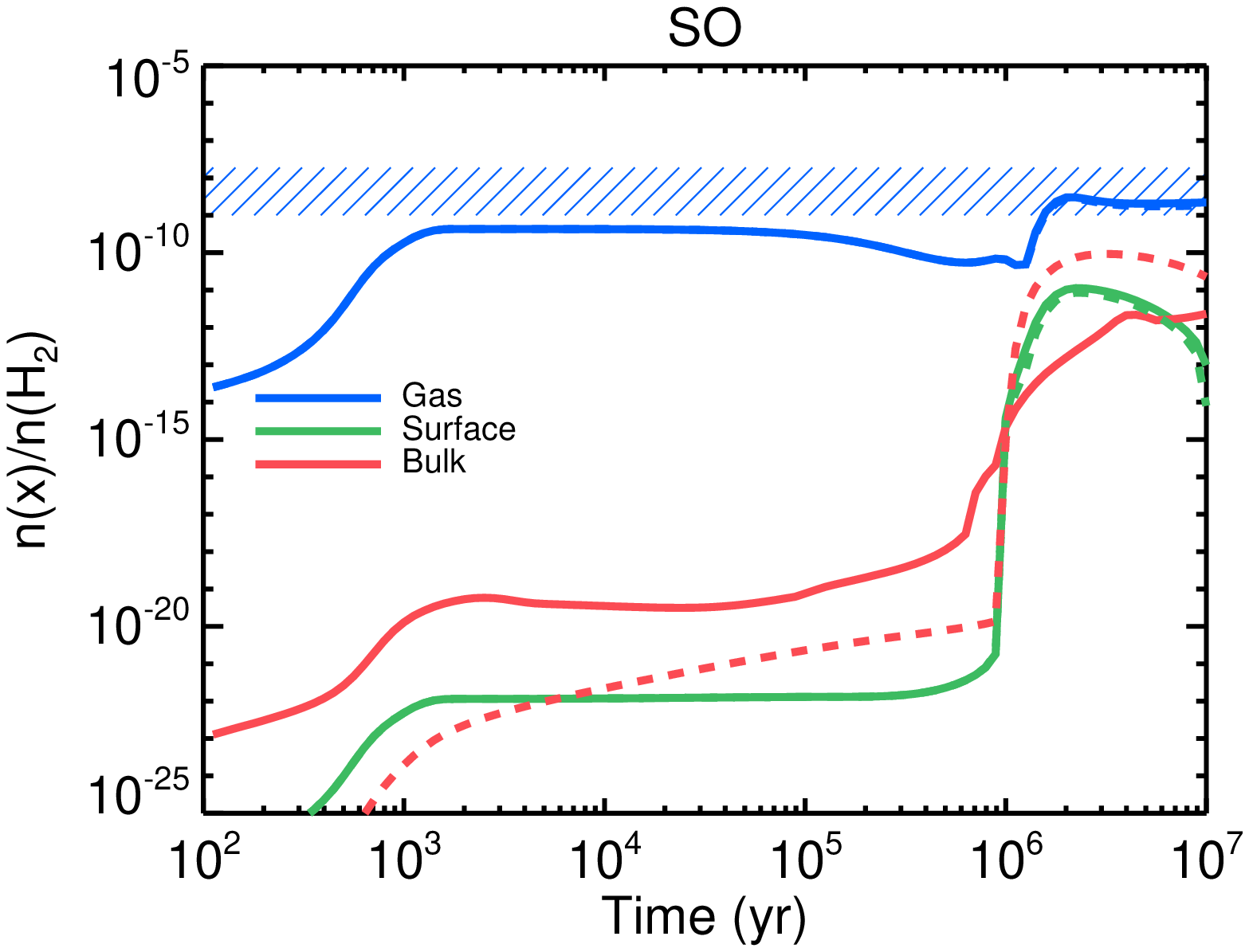}{0.5\textwidth}{}
          \fig{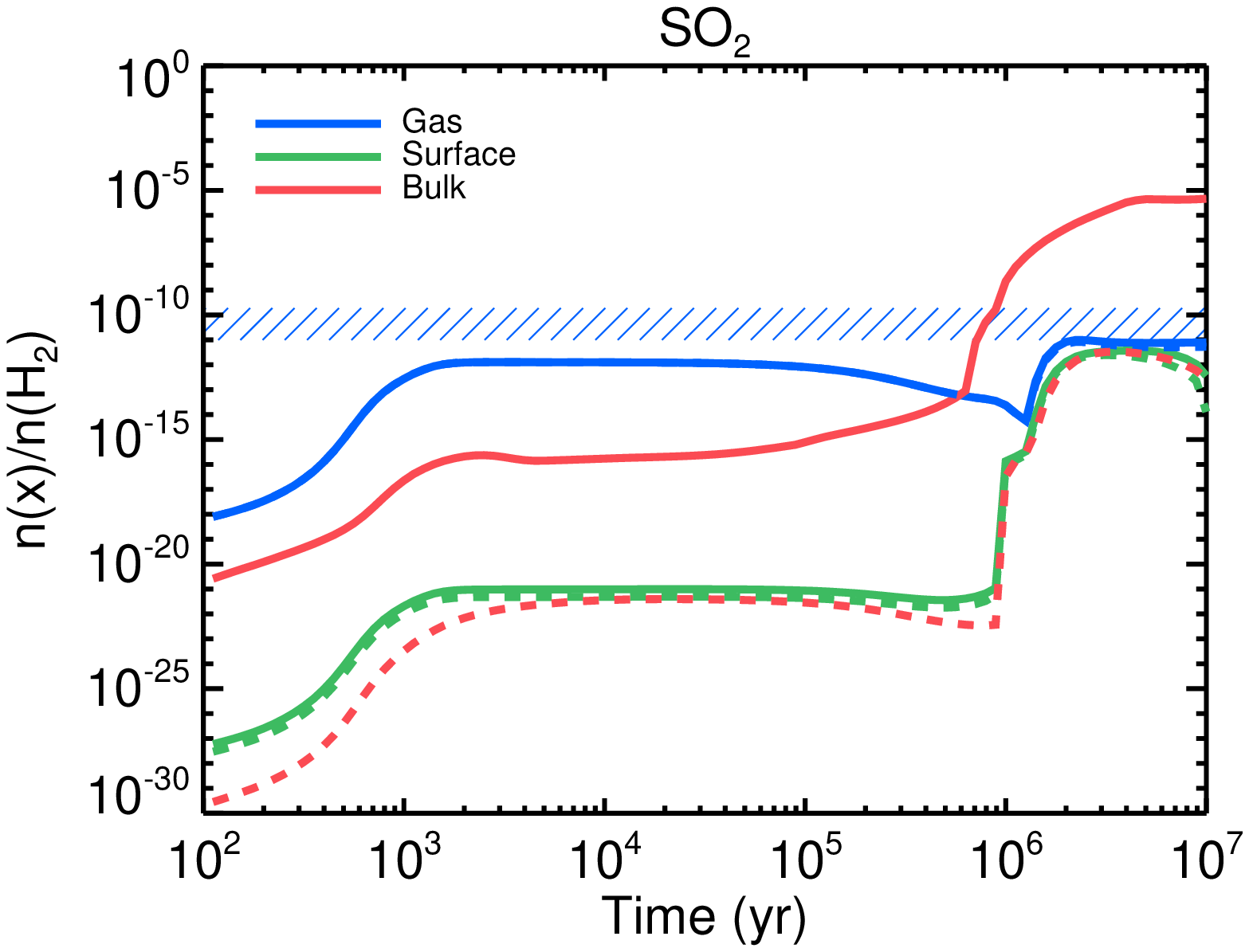}{0.5\textwidth}{}
}
\gridline{
          \fig{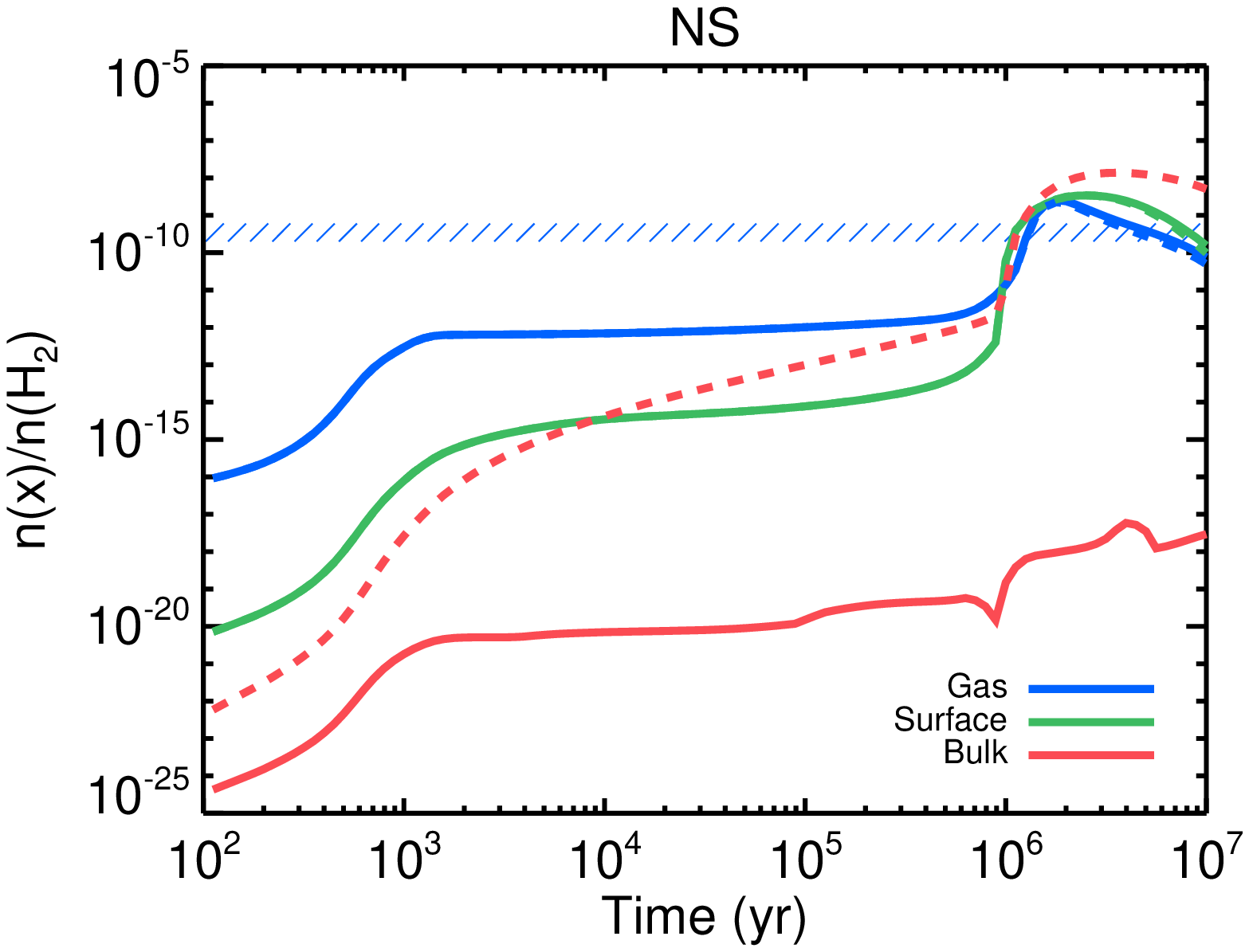}{0.5\textwidth}{}
          \fig{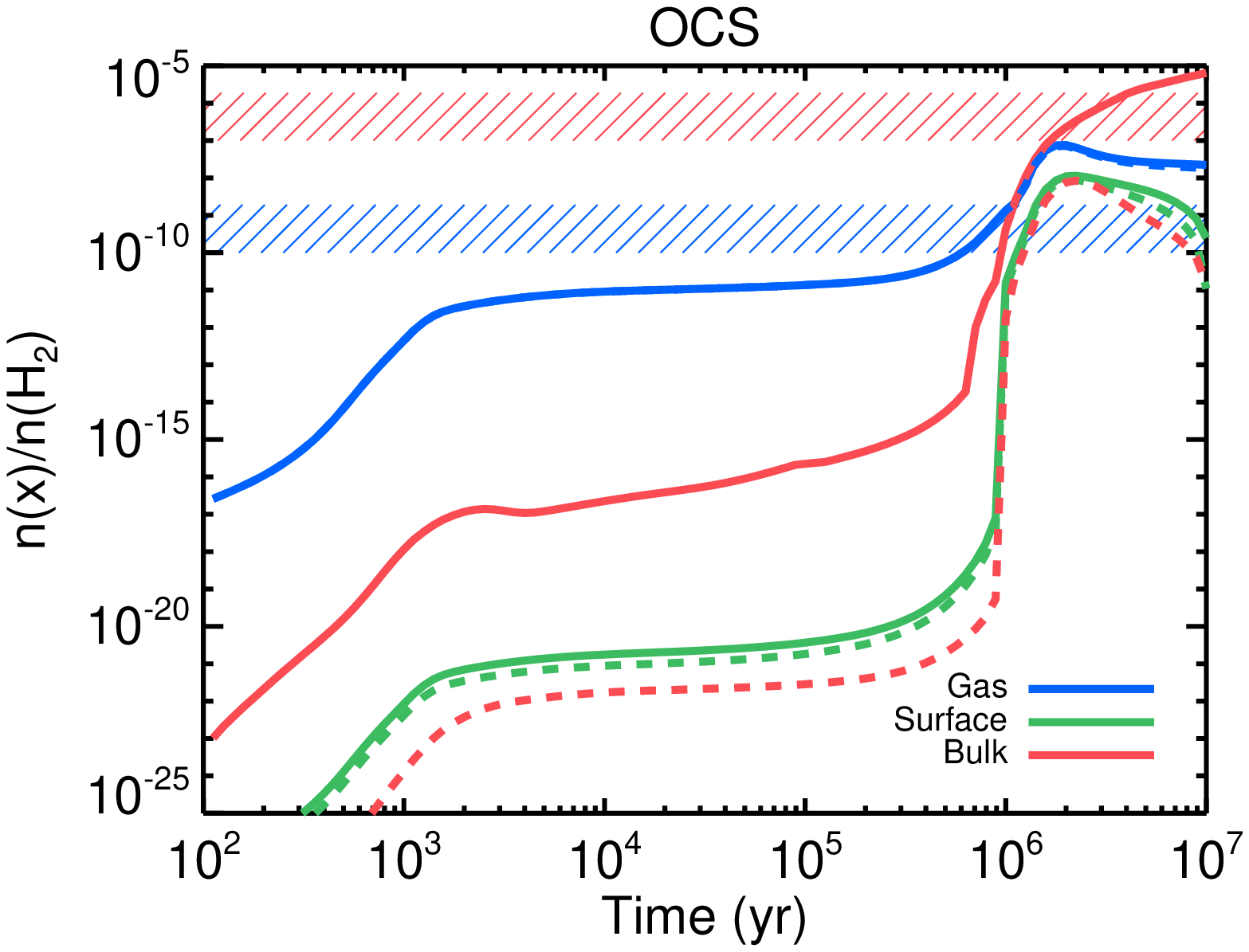}{0.5\textwidth}{}
}
\caption{ Calculated abundances of S, CS, \ce{SO}, \ce{SO2}, \ce{NS}, and \ce{OCS} in models A (dashed line) and B (solid line). Gas-phase observational abundances for dense clouds, where available, are represented by horizontal blue hatched bars (see \citet{laas_modeling_2019} and references therein.).
\label{fig:block2}}
\end{figure*}

The results of our simulations for models A and B are presented in Figs. \ref{fig:block1} - \ref{fig:allotropes}. In Figs. \ref{fig:block1} and \ref{fig:block2}, the gas (blue), surface (green), and bulk (red) abundances with respect to molecular hydrogen of a number of sulfur-bearing species are shown. Where available, gas-phase abundances from dense cloud observations are represented by blue hatched bars, with values taken from Table 4 of \citet{laas_modeling_2019}. In all figures, results from Model A are depicted with dashed lines, and those from Model B by solid lines. We note that, unless otherwise stated, reactions referred to here describe processes occuring within the bulk of the ice.

Table \ref{tab:comet} compares the abundances of a number of sulfur-bearing species in the comet nucleus (inferred from analysis of gas-phase coma material measured by \textit{Rosetta}), with the total ice abundance (surface + bulk) in Models A and B. Calculated abundances listed in Table \ref{tab:comet} are those at model times of $\sim2\times10^6$ yr (corresponding to an ice ca. $100$ monolayers thick) since, as we show below, this is the point in our simulations of best agreement with existing observational data. Our results at this time, therefore, may provide an interesting point of comparison between ISM values and those from much older Solar-system objects. In addition to the obvious differences in age, however, we stress that care should also be taken in interpreting the data in Table \ref{tab:comet} since the cometary abundance of some species, e.g. \ce{S2} and \ce{SO}, may have been enhanced due to the fragmentation of some larger parent species either within the mass spectrometer on board \textit{Rosetta}, or directly within the nucleus of 67P/C-G.

Of the species whose abundances are depicted in Figs. \ref{fig:block1} and \ref{fig:block2}, eight have been detected in cold dense clouds, namely \ce{H2S}, \ce{C2S}, \ce{H2CS}, \ce{CS}, \ce{SO}, \ce{SO2}, \ce{NS}, and \ce{OCS} (see \citet{laas_modeling_2019} and references therein). These two figures show that our models agreeably reproduce the abundances of most of these observed species at around $1-2\times10^6$ yr - a reasonable time for such dense clouds - with the exception of CS and, to a lesser degree, \ce{H2S}, both of which are overproduced in the gas. One can see, moreover, that while the novel processes included in Model B have profound effects on the bulk abundances of most species shown, as expected, the two models predict generally similar gas and surface abundances. 

\subsection{Effect of novel bulk processes}

\subsubsection{Radicals}

Not surprisingly, based on our previous results described in \citet{shingledecker_simulating_2019}, one clear effect of our non-diffusive mechanism is a significant reduction in the abundances of radicals in the bulk, as can be seen by, e.g. NS, CS, and S in Fig. \ref{fig:block2}. 
Models often predict large abundances of these reactive species in interstellar ices. It may be, though, that such results represent worrying departures from physical realism, a supposition supported somewhat surprisingly by the results of \citet{greenberg_exploding_1973} in their well-known work describing the dramatic phenomenon they called ``grain explosions.'' There, Greenberg and Yencha speculated that such explosions were driven by the exothermic reactions of radicals trapped in the ice. Critically, though, they note that in reality, the collective concentration of these reactive species should not exceed $\sim1\%$ due to the rapidity with which they react both with themselves and their neighbors. This implies that S, HS, and other reactive species can most likely be ruled out as significant reservoirs of sulfur in dust-grain ice mantles.

\subsubsection{\ce{H2S}}

Hydrogen sulfide, \ce{H2S}, is a species often predicted to be a major reservoir of sulfur in dense regions \citep{esplugues_modelling_2014,holdship_h2s_2016,danilovich_sulphur-bearing_2017,vidal_reservoir_2017}, since it forms readily from sulfur atoms adsorbed onto grain surfaces via

 \begin{equation}
     \ce{H + S -> HS}
     \label{h2smake1}
 \end{equation}
 
 \noindent
 followed by
 
 \begin{equation}
    \ce{H + HS -> H2S}. 
    \label{h2smake2}
 \end{equation}

\noindent
Thus, as dust-grain ice mantles grow, \ce{H2S} is trapped in the bulk where, being resistant to further reaction to hydrogen due to the large reaction barrier, it remains by far the most abundant sulfur-bearing species on the grain. 

However, there are two main problems with the hypothesis that hydrogen sulfide is the primary sulfur reservoir in dust-grain ice mantles. First, in addition to driving the formation of \ce{H2S}, atomic hydrogen can also efficiently destroy hydrogen sulfide even at low temperatures via tunneling, as was shown recently by \citet{lamberts_tunneling_2017}. The second, major flaw is that it is apparently not sufficiently abundant in dust-grain ice mantles to be detected \citep{smith_search_1991,boogert_observations_2015}.

As shown in Fig. \ref{fig:block1}, in agreement with past modeling results, Model A likewise predicts that a substantial fraction of the total sulfur is locked in \ce{H2S} at late times. Strikingly, though, in Model B, bulk abundances of \ce{H2S} are reduced by $\sim5-6$ orders of magnitude compared with those in Models A. 
This reduction in bulk abundance is mainly due to two factors at play in Model B; namely, (1) the increased destruction of \ce{H2S} in the bulk by cosmic rays, as well as reactions with atoms and radicals, especially OH, and, (2) the fact that in Model B, bulk reactions are not dominated solely by those involving light, mobile species such as atomic hydrogen, and thus, that the HS produced in reaction \eqref{r1} does not quickly reform \ce{H2S} via \eqref{h2smake2}. 

The contrasting results predicted by Models A and B serve as a good illustration of the kinds of chemistry that characterize reactions in the bulk, depending on the assumptions made regarding the underlying mechanisms. Specifically, in the traditional diffusive approach, bulk chemistry is dominated by reactions involving atomic hydrogen, given its abundance, reactivity, and high mobility, whereas our new non-diffusive approach allows other radicals, such as HS - which would otherwise either build up or form \ce{H2S} - to play a more active role. 

From Table \ref{tab:comet}, one can see that both models are within an order of magnitude of the \ce{H2S} abundance measured by \textit{Rosetta}, though with the Model A result somewhat closer to the cometary value. Further comparison between the average [\ce{H2S}]/[\ce{H2O}] upper limit 
derived by \citet{smith_search_1991} of 1.5\% in dense interstellar clouds, and our calculated value at 2 Myr confirms that the Model A result of 3\% is likely too large, with the Model B value of 0.4\% being in better agreement with the observational data in this regard.

\subsubsection{\ce{OCS}}

Carbonyl sulfide, OCS, the only sulfur-bearing species definitively detected in interstellar ices to date \citep{palumbo_solid_1997,aikawa_akari_2012}, serves as an important means by which we can quantitatively compare how accurately our models are simulating the real chemistry of dust-grain ice mantles. From Fig. \ref{fig:block2}, one can see that the bulk abundance of OCS is substantially higher in Model B than Model A. This increase is driven mainly by the grain reaction

\begin{equation}
    \ce{S + CO -> OCS},
\end{equation}

\noindent
which occurs with only a negligible rate in Model A due to the slow diffusion of the reactants at 10 K.

Moreover, when we compare our data with the available observational findings, it becomes clear that Model B provides a much better match with existing empirical data. Specifically, Model B reproduces the solid OCS relative abundance of $\sim10^{-6}$, as well as the ice/gas abundance ratio of $\sim10^3$ observed by \citet{aikawa_akari_2012}. Further comparison with [OCS]/[CO] abundance ratios measured by \citet{palumbo_solid_1997}, listed in Table \ref{tab:ocs}, shows again that the results of Model B are in better agreement with previous observational results. One draws a similar conclusion from a comparison of our simulation results and data from the \textit{Rosetta} mission which, again, more closely match calculated OCS ice abundances from Model B.

\begin{deluxetable}{lD}[bt!]
\tablecaption{Observed and calculated [OCS]/[CO] abundance ratios in interstellar dust-grain ice mantles. \label{tab:ocs}}
\tablewidth{0pt}
\tablehead{
\colhead{Source} & \twocolhead{Abundance (\%)} 
}
\startdata
\multicolumn{3}{c}{Observed} \\
\decimals
W33A           & $5.0\times10^{0}$ \\
AFGL 989       & $8.0\times10^{-1}$ \\
Mon R2 IRS2    & $6.5\times10^{-1}$  \\
AFGL 961E      & $<1.0\times10^{0}$  \\
AFGL 490       & $<2.6\times10^{0}$  \\
NGC 2024 IRS 2 & $<8.0\times10^{-1}$ \\
OMC 2 IRS 3    & $<1.6\times10^{0}$  \\
Elias 16       & $<8.0\times10^{0}$ \\ \hline
\multicolumn{3}{c}{Calculated} \\
Model A        & $1\times10^{-2}$ \\
Model B        & $7\times10^{-1}$ \\
\enddata
\tablecomments{Observational values are taken from \citet{palumbo_solid_1997}, and calculated values are those at $t\approx2\times10^6$ yr.}
\end{deluxetable}

\subsubsection{\ce{SO2}}

Although OCS remains the only definitely-detected sulfur-bearing species in interstellar ices, there is also tentative evidence for the presence of sulfur dioxide, \ce{SO2}, based on observations by \citet{zasowski_spitzer_2009} and \citet{boogert_infrared_1997}, who report values of [\ce{SO2}]/[\ce{H2O}]$\approx0.5\%$ and $\approx0.1-1\%$, respectively. That \ce{SO2} is indeed an important component of interstellar ices is perhaps hinted at by the ubiquity of sulfur dioxide frost on the Jovian moon, Io - the surface of which is dominated by abundant sulfur-bearing molecules, particularly \ce{S8} \citep{carlson_ios_2007}. Thus, \ce{SO2} represents another key species useful in determining the relative accuracy of our simulations. 

In this regard, as shown in Fig. \ref{fig:block2}, Model B again performs better: predicting a bulk abundance $\sim$5 orders of magnitude greater than Model A. As with OCS, this increase is due to eliminating the reliance on thermal diffusion within the ice mantle, thereby allowing sulfur dioxide to form \textit{in situ} efficiently via 

\begin{equation}
    \ce{SO + OH -> SO2 + H},
\end{equation}

\noindent
unlike in Model A, where \ce{SO2} on grains is predominantly the result of accretion from the gas. 

Comparing our calculated sulfur dioxide abundance with those from \citet{zasowski_spitzer_2009} and \citet{boogert_infrared_1997}, we note that Model B reproduces their reported [\ce{SO2}]/[\ce{H2O}]$\approx0.5\%$ at ca. 2 Myr, with the abundance in Model A at that time being $\approx10^{-6}\%$. Interestingly, \textit{Rosetta} detected an \ce{SO2} abundance of [\ce{SO2}]/[\ce{H2O}]$\approx0.1\%$, remarkably similar to what has been reported in these tentative detections. The \ce{SO2} ice abundance predicted by Model B is therefore in agreement with both cometary results as well as the existing ISM values, with Model A likewise underproducing \ce{SO2} in both cases by $\sim$5 orders of magnitude. 

\subsubsection{Sulfur Allotropes}

\begin{figure*}[t!]
%    \hspace{-1cm}
    \centering
    \includegraphics[width=0.6\paperwidth]{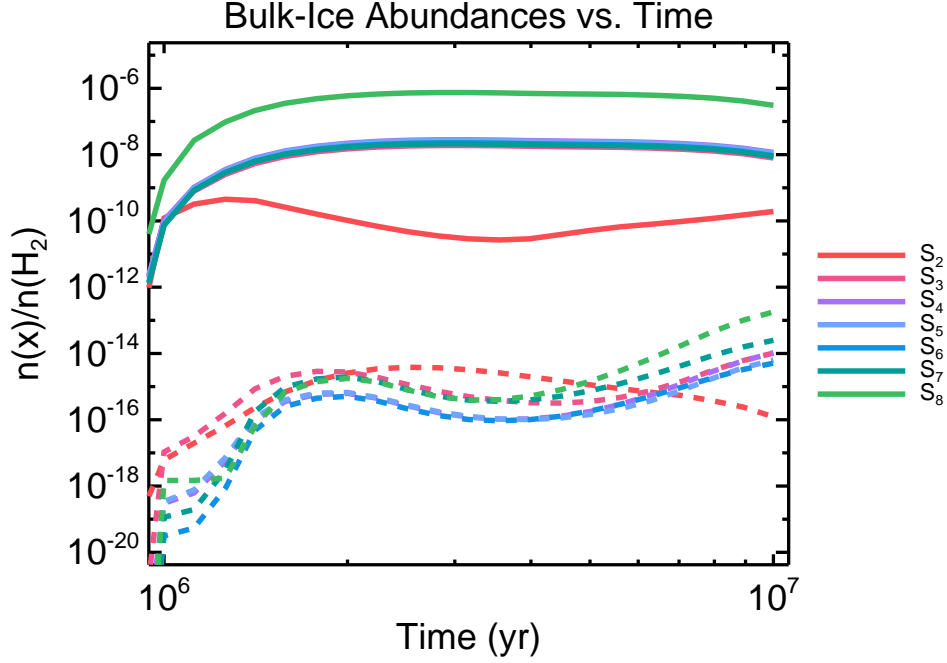}
    \caption{Abundances of sulfur allotropes ($\mathrm{S}_n$, $n\in [2,8]$) within the bulk in models A (dashed line) and B (solid line).}
    \label{fig:allotropes}
\end{figure*}

Over 30 different allotropes of sulfur are currently known - more than any other element \citep{steudel_solid_2003}. Of these,  \ce{S8} - also called octasulfur or simply ``elemental sulfur'' - is the most stable, and by far the most abundant in nature. Octasulfur, one form of which has the bright-yellow color historically associated with deposits of the pure element on Earth \citep{steudel_homocyclic_1982}, can also be seen on other Solar-system bodies, such as Io, which is thought to have a surface rich in \ce{S8} based, in part, on its coloration and the observation of \ce{S8} features in UV-Vis Solar reflectance spectra \citep{carlson_ios_2007}. 

One of the most interesting results of this study is the combined effect of radiation chemistry and non-diffusive bulk reactions on the abundances of $\mathrm{S}_n$ ($n\in[2,8]$). As can be seen from Fig. \ref{fig:allotropes}, the increased efficiency of bulk chemistry in Model B results in substantially higher abundances for the all sulfur allotropes in our network, in particular \ce{S8}. 

The formation of these pure-sulfur species begins with \ce{S2}, which in Model B occurs mainly via \citep{mihelcic_esr-spektroskopische_1970}

\begin{equation}
    \ce{S + HS -> S2 + H}.
\end{equation}

\noindent
Once produced, disulfur can then dimerize to form \ce{S4} which can, for example, either react with S to form \ce{S5} or with \ce{S2} to form \ce{S6}. Conversely, in Model A, \ce{S2} in the mantle comes either from the accretion of gas-phase \ce{S2} or, to a lesser degree, the surface reaction \citep{sendt_chemical_2002}

\begin{equation}
    \ce{H + S2H -> S2 + H2}.
\end{equation}

\noindent
The results of Model B are in agreement with previous experimental studies on interstellar ice analogues, which show that \ce{S8} can form readily at low temperatures in both UV-  \citep{chen_formation_2015,jimenez-escobar_sulfur_2011} and proton- \citep{garozzo_fate_2010} irradiated ices; however, these are the first simulations - to the best of our knowledge - which show that the formation of sulfur allotropes can indeed be efficient in real astrophysical environments.

Intriguingly, the presence of elemental sulfur in cometary nuclei was suggested recently by the detection of \ce{S2}, \ce{S3}, and \ce{S4} during the \textit{Rosetta} mission, which \citet{calmonte_sulphur-bearing_2016} noted were likely formed from a compound such as \ce{S8}. A comparison of our calculated abundances with \textit{Rosetta} measurements of \ce{S2} shows, again, that Model B yields values closer to the available data.

\subsubsection{Sulfur Reservoir}

\begin{figure*}
    \centering
    \gridline{
          \fig{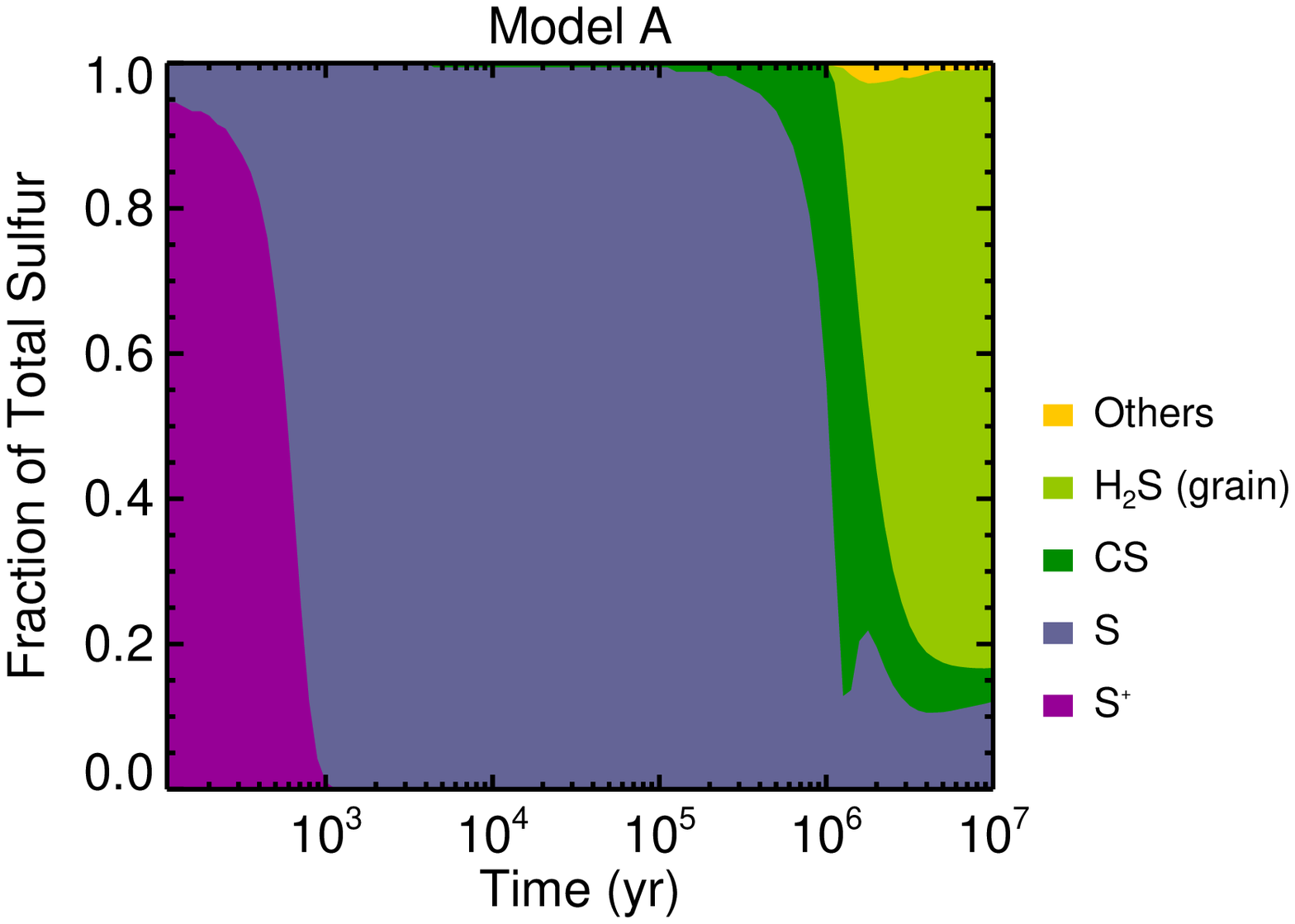}{0.5\textwidth}{}
          \fig{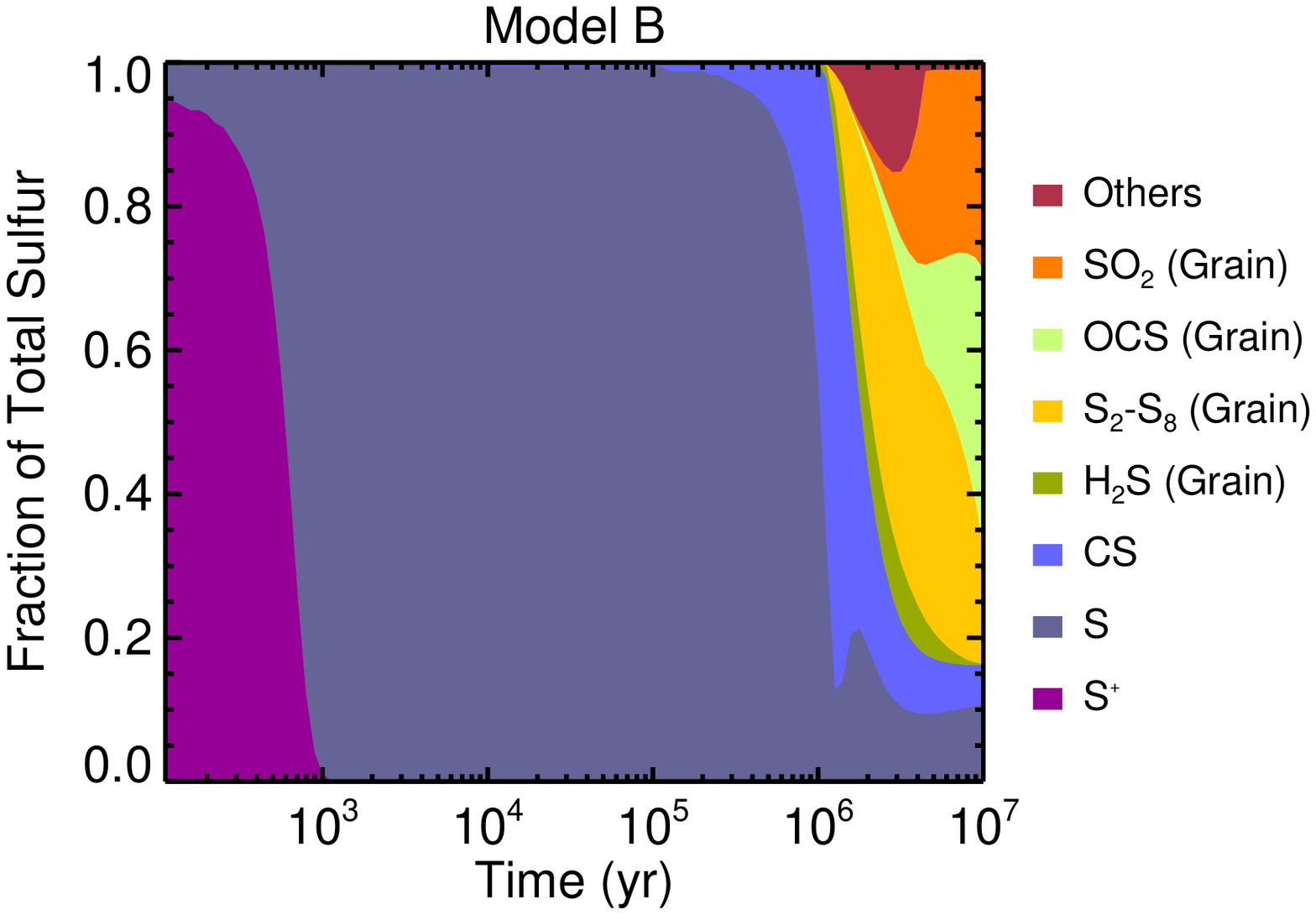}{0.5\textwidth}{}
          }
    \caption{Main sulfur-bearing species in Model A (left) and Model B (right).}
    \label{fig:wheresulfur}
\end{figure*}

Shown in Fig. \ref{fig:wheresulfur} are the dominant sulfur-bearing species as a function of time in Models A and B. At times before ca. 1 Myr, both models predict similar progressions of the dominant sulfur-bearing species from the initial \ce{S+}, to neutral atomic sulfur, and finally CS. After 1 Myr, Models A and B likewise show similar degrees of depletion of gas-phase sulfur onto grains, though differences in how bulk chemistry is treated in the two models lead to strikingly dissimilar predicted abundances for sulfur-bearing species in the ice.

Model A - our fiducial simulation - predicts an ice in which nearly all sulfur exists as \ce{H2S}. This high hydrogen sulfide abundance has been characteristic of previous model results \citep{vidal_reservoir_2017,holdship_h2s_2016}, and is the natural outcome of a diffusive surface and bulk chemistry dominated by reactions with atomic hydrogen, given its reactivity, mobility, and abundance \citep{jimenez-escobar_sulfur_2011}. 
One exception was the recent study by \citet{laas_modeling_2019}, who found SO and OCS to be more abundant than \ce{H2S} on the grain. Though our network is based on the one presented in that work, their use of a two-phase model means that diffusion and desorption are more efficient than in three-phase codes such as the one we use in this study - resulting in a greater similarity between the results of \citet{laas_modeling_2019} and those of Model B.

One alternative candidate for the long-sought sulfur reservoir is suggested by our Model B results. In Fig. \ref{fig:wheresulfur}, one can see three major classes of sulfur-bearing species on grains at late times, namely, OCS, \ce{SO2}, and the allotropes. It has been speculated previously by \citet{palumbo_solid_1997} and \citet{garozzo_fate_2010} that the presence of large amounts of \ce{S8} (and related species) could resolve the missing-sulfur problem. Unfortunately, \ce{S8} has no strong IR-active modes and it is therefore very difficult to observationally constrain its abundance in real dust-grain ice mantles \citep{palumbo_solid_1997}. Even observations of post-shocked material, as done recently by \citet{holdship_sulfur_2019}, also might not clearly reveal the presence of abundant sulfur allotropes, because of their refractory nature or the difficulty of observing even their dissociation products, e.g. \ce{S4}. Nevertheless, it was recently inferred by \citet{kama_abundant_2019}, that much of the sulfur in protoplanetary disks is locked away in such a refractory compound with physical characteristics consistent with those of the sulfur allotropes.

Shown in Fig. 9 of Appendix B are the corresponding plots showing the major sulfur-bearing species predicted by Models C - E. Fig. S4 shows that, of the processes listed in Table \ref{tab:models}, non-diffusive radical reactions have the single largest effect on the resulting abundances of bulk species. By comparison, cosmic-ray-driven chemistry on its own has a smaller effect at low temperatures, since the resulting thermal radiolysis products build up in the ice due to their low diffusion-based reaction rates. As we first showed in \citet{shingledecker_simulating_2019}, our Model B results further illustrate the importance of non-diffusive bulk reactions for fully and accurately simulating solid-phase radiation chemistry. Critically though, Figs. \ref{fig:wheresulfur} and S4, show that the large sulfur allotrope abundances predicted in Model B require \textit{both} of these mechanisms.

\FloatBarrier
\section{Conclusions} \label{sec:conc}

In this work, we have examined the effects of novel modifications of astrochemical models on the abundances of sulfur-bearing species in cold cores. Specifically, we have examined the effects of, (a) a cosmic ray-driven radiation chemistry, and (b) fast \textit{non-diffusive} bulk reactions for radicals and reactive species.

Our main results are the following: 

\begin{enumerate}
    \item The inclusion of (a) and (b) in three-phase models results in increased bulk abundances of OCS and \ce{SO2}, leading to good agreement between the calculated and observed abundance ratios for these species in interstellar ices. 
    \item Moreover, (a) and (b) improved the overall agreement between the results of our simulations and the abundances of sulfur-bearing species measured by the \textit{Rosetta} mission, shown in Table \ref{tab:comet}  \citep{calmonte_sulphur-bearing_2016}. 
    \item Finally, the inclusion of (a) and (b) greatly increases the bulk abundance of sulfur allotropes, particularly elemental sulfur (\ce{S8}). 
\end{enumerate}

The non-diffusive mechanism we have used for treating the reaction of radicals in the bulk has been tested in a previous work and shown to yield far better agreement with well-constrained experiments than current approaches relying on thermal diffusion of one kind or another \citep{shingledecker_simulating_2019}. This work is a first attempt to apply such insights to models of the ISM, and thus, Model B may represent the most realistic simulation of interstellar ice-mantle chemistry to date. Nevertheless, much work can and should still be done to improve both our chemical network as well as the physical processes simulated by the code itself. 

One promising method for improving solid-phase chemical networks is to attempt to reproduce well-constrained experiments, as we have done previously with ion-irradiated pure \ce{O2} and \ce{H2O} ices \citep{shingledecker_simulating_2019}. In many current laboratory studies, however, the abundances of only a small number of species can be tracked during the course of the experiment using traditional techniques such as FTIR. Unfortunately, the physical properties of elemental sulfur that make it difficult to observe in the ISM make measuring its abundance in experimental ices similarly difficult. Nevertheless, \ce{S8} can be detected using Raman spectroscopy \citep{anderson_low_1969}, which may be the only practical way to estimate whether, or to what degree, such species form in sulfur-containing ice mantles under interstellar conditions. 

Regarding additional physical effects resulting from bombardment by energetic particles, perhaps the most astrochemically important of these are related to the desorption of ice-mantle species. Some, such as desorption stimulated by grain heating, are currently considered in a preliminary way \citep{hasegawa_new_1993}, and could be improved with more accurate estimates related to the average amount of heating per cosmic ray as well as the rate at which such heat propagates through the ice mantle - including as a function of grain size, which \citet{zhao_effect_2018} found could have significant effects on resulting gas-phase abundances. In addition, though, energetic particle bombardment is known to drive desorption via a number of \textit{non-thermal} mechanisms including sputtering \citep{burkhardt_modeling_2019} and impulsive spot heating \citep{ivlev_impulsive_2015}, and attempts should be made to include such processes in future models. 

C.N.S. thanks the Alexander von Humboldt Foundation for their generous support and both K. Altwegg and W.F. Thi for stimulating discussions during the course of this project. T.L. is grateful for support from the Netherlands Organisation for Scientific Research (NWO) via a VENI fellowship (722.017.008). E. H. thanks the National Science Foundation for support of his program in astrochemistry. The work by A.V. is supported by Russian Science Foundation via the Project 18-12-00351. A.V. is the head of Partner Group of the Max Planck Institute for Extraterrestrial Physics, Garching, at the Ural Federal University, Ekaterinburg, Russia. 

\software{MONACO \citep{vasyunin_formation_2017}}

\appendix

\section{Chemical Network}

Expanded sulfur network used here. Here, R1 \& 2 are the reactants, while P1 - P5 represent the products. The parameters $\alpha$, $\beta$, and $\gamma$ are used in calculating rate coefficients. For specific formulae used, see \citet{laas_modeling_2019} and \citet{semenov_chemistry_2010}. 

\begin{deluxetable}{ccccccccccc}[h]
\tablecaption{Chemical network used in this work, based on that of \citet{laas_modeling_2019}. \label{tab:network}}
\tablehead{
\colhead{\#} &
\colhead{R1} & \colhead{R2} & 
\colhead{P1} & \colhead{P2} & \colhead{P3} & \colhead{P4} & \colhead{P5} &
\colhead{$\alpha$} & \colhead{$\beta$} & \colhead{$\gamma$} \\
}
\startdata
1  & C$^+$   & G$^-$ & C     & G$^0$  &         &       &      & 1.00E+00 & 0.00 & 0.0 \\
2  & Fe$^+$  & G$^-$ & Fe    & G$^0$  &         &       &      & 1.00E+00 & 0.00 & 0.0 \\ 
3  & G$^0$   & e$^-$ & G$^-$ &        &         &       &      & 1.00E+00 & 0.00 & 0.0 \\
4  & H$^+$   & G$^-$ & H     & G$^0$  &         &       &      & 1.00E+00 & 0.00 & 0.0 \\
5  & H$_3^+$ & G$^-$ & H$_2$ & H      & G$^0$   &       &      & 1.00E+00 & 0.00 & 0.0 \\
\enddata
\tablecomments{Table \ref{tab:network} is published in its entirety in the machine readable format. A portion is shown here for guidance regarding its form and content.}
\end{deluxetable}

\section{Results from Models C - E} 

In order to disentangle the effects of the various novel features we have included in our model, we have run additional simulations, detailed in Table 4. The results of these models (C-E) are shown here for comparison.

\begin{figure}
\gridline{
          \fig{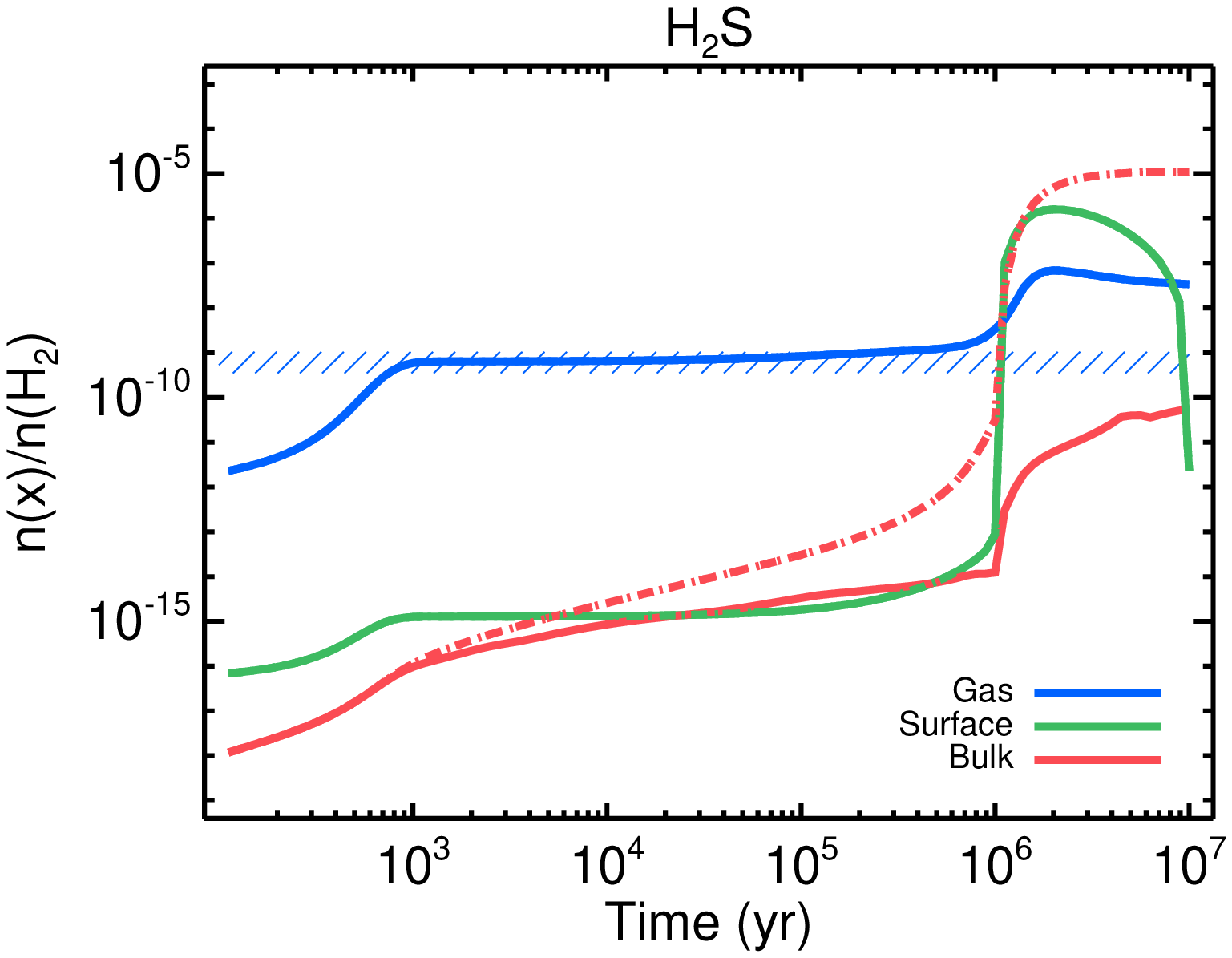}{0.5\textwidth}{}
          \fig{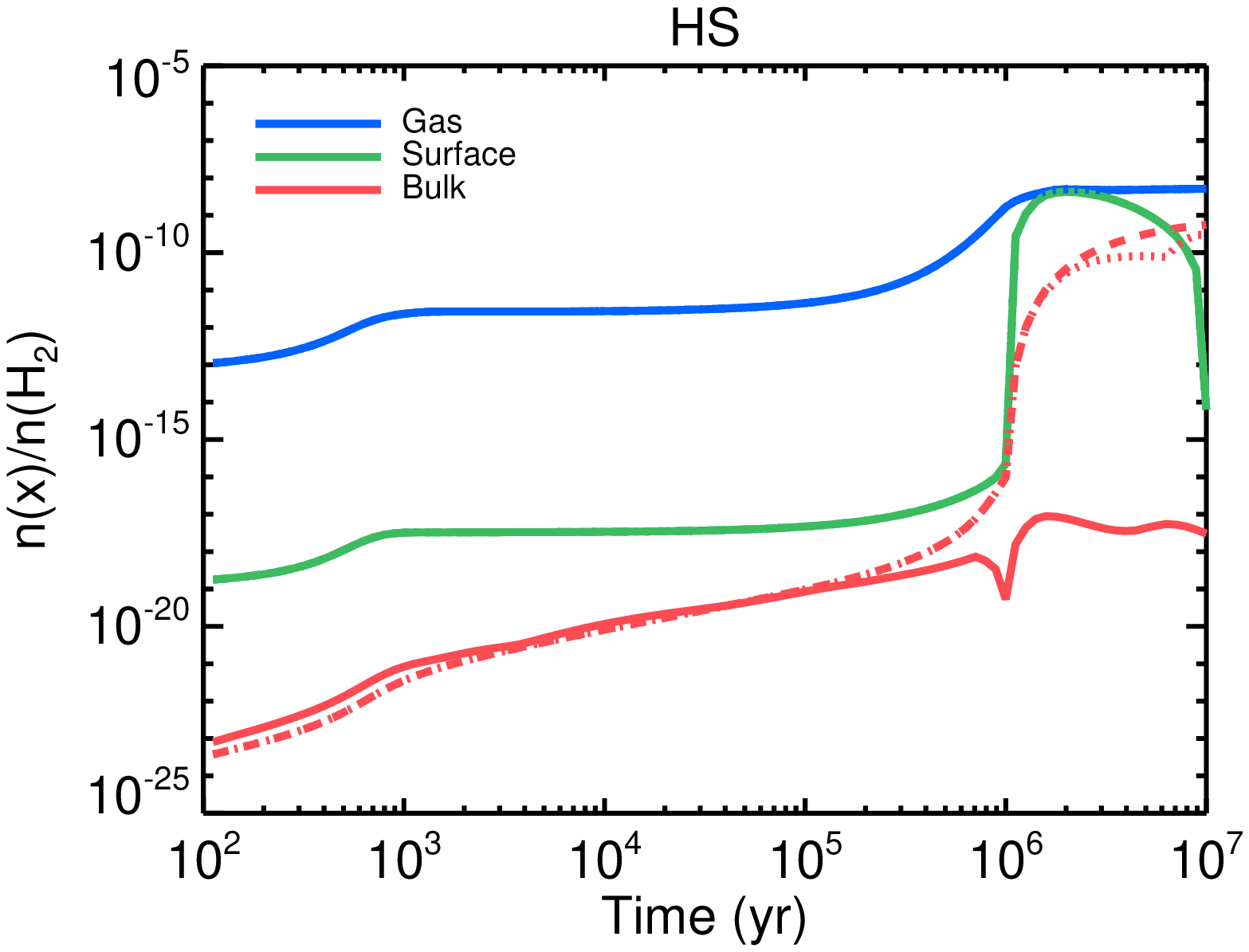}{0.5\textwidth}{}
          }
\gridline{
          \fig{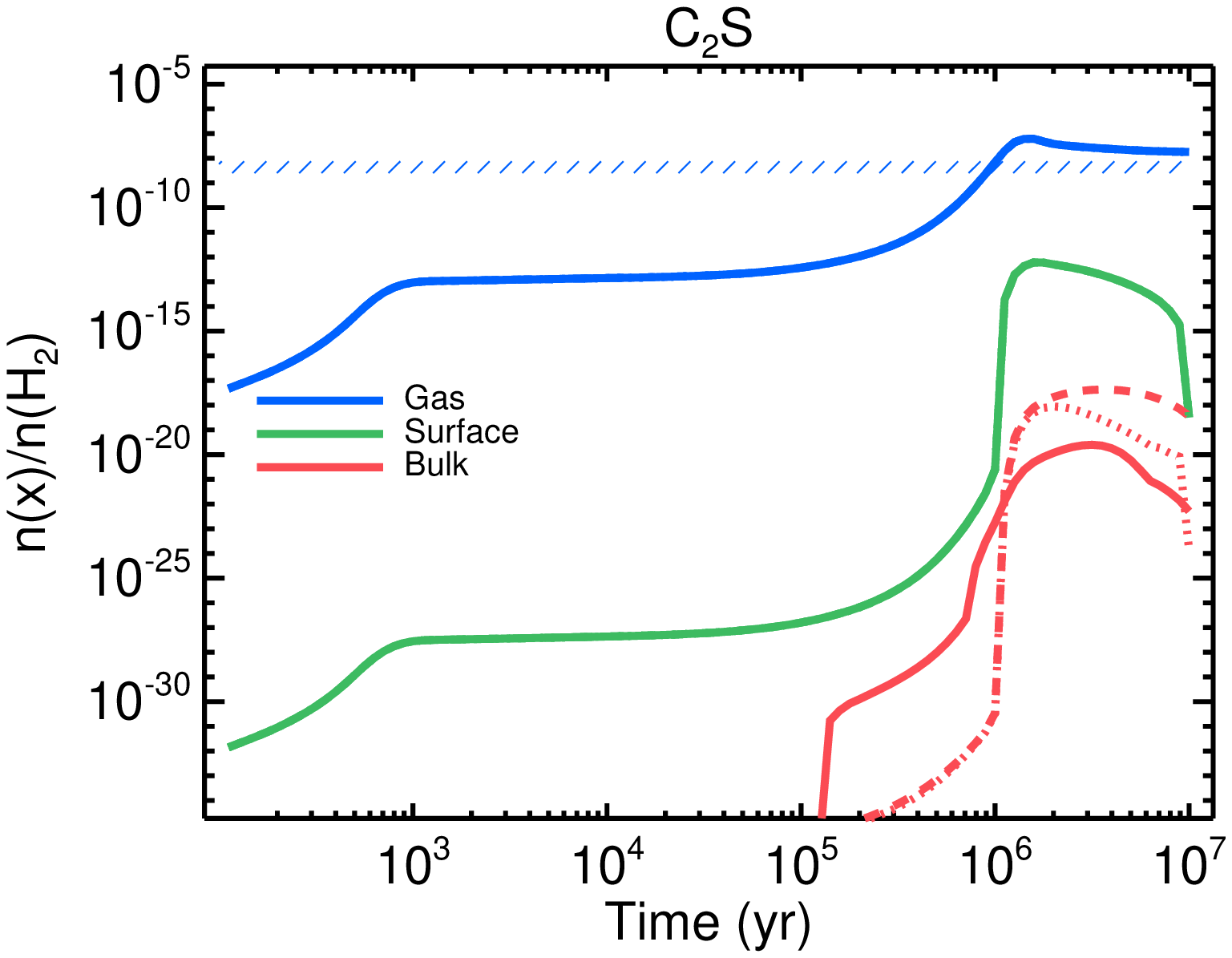}{0.5\textwidth}{}
          \fig{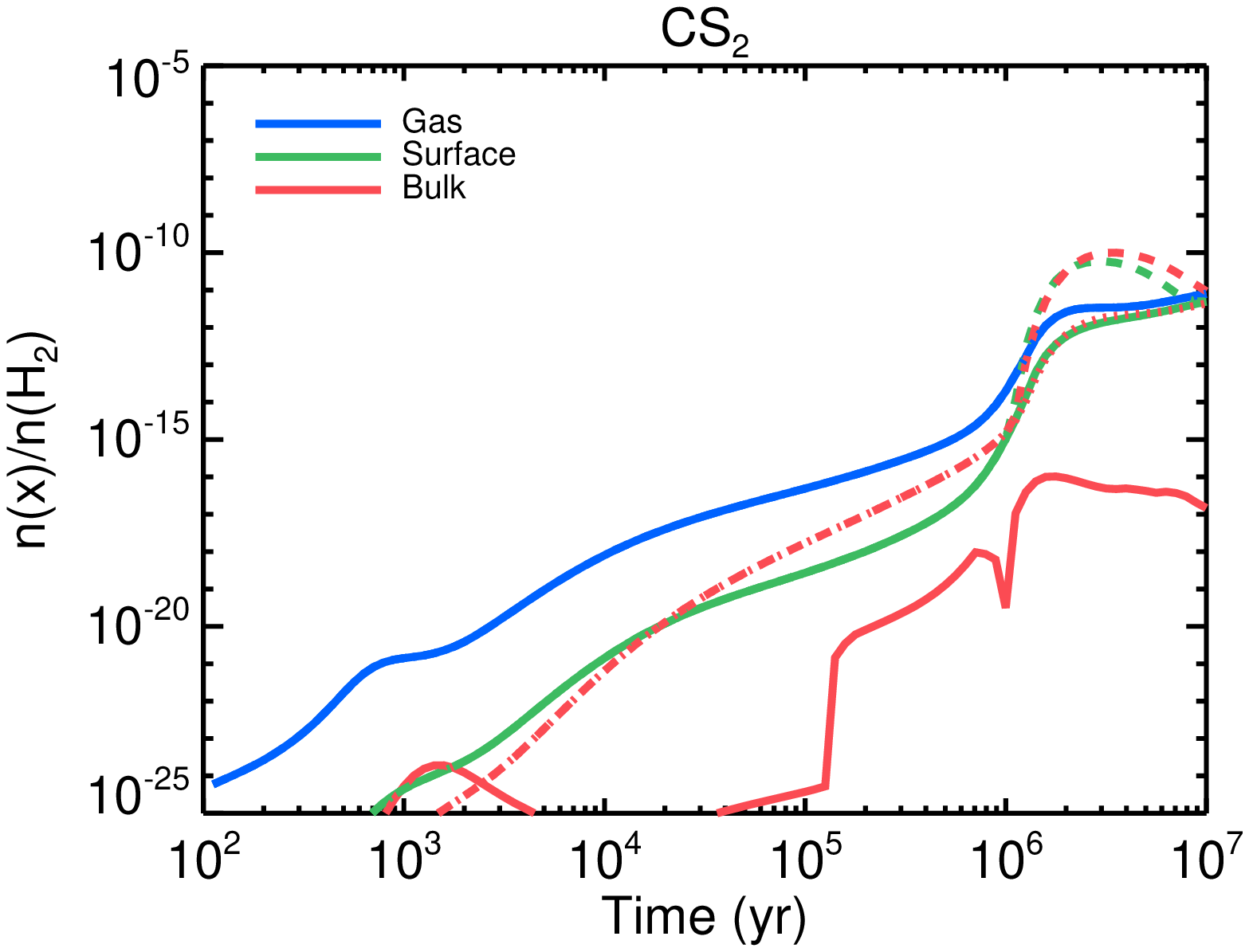}{0.5\textwidth}{}
}
\gridline{
          \fig{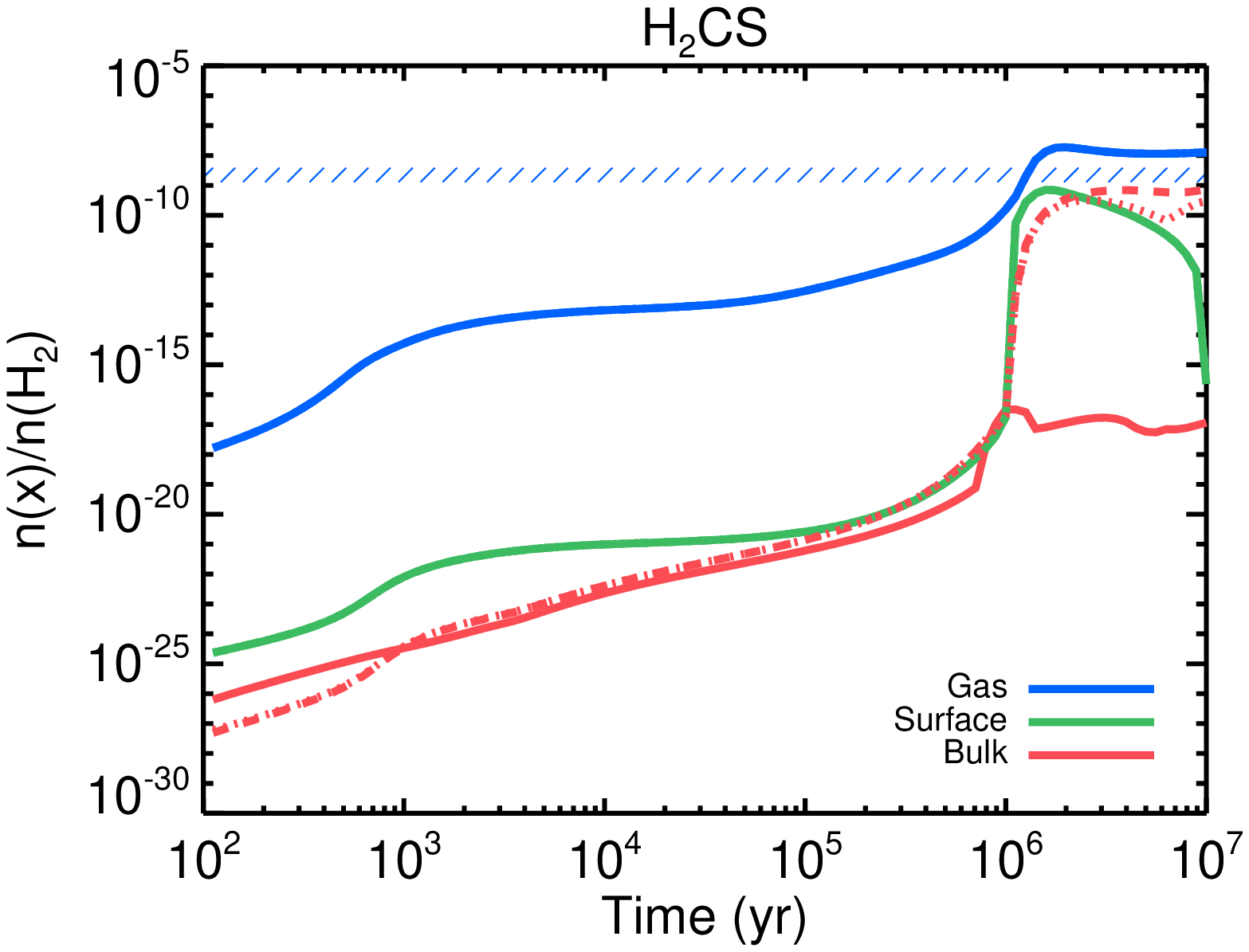}{0.5\textwidth}{}
          \fig{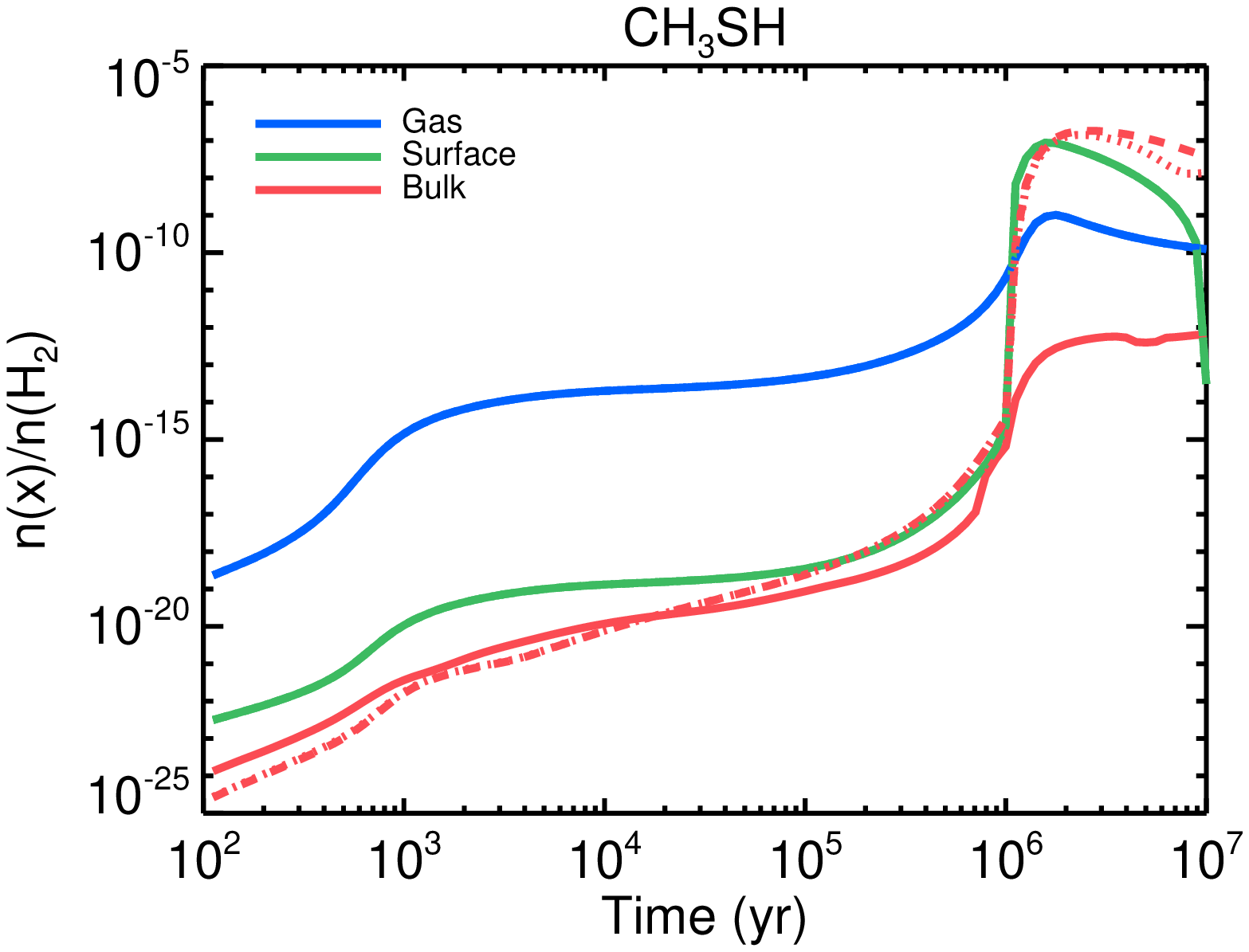}{0.5\textwidth}{}
}
\caption{ Calculated abundances of \ce{H2S}, \ce{HS}, \ce{C2S}, \ce{CS2},  \ce{H2CS}, and \ce{CH3SH} in models C (solid line),  D (dashed line), and E (dotted line). Gas-phase observational abundances for dense clouds, where available, are represented by horizontal blue hatched bars (see \citet{laas_modeling_2019} and references therein.).
\label{fig:supblock1}}
\end{figure}

\begin{figure}
\gridline{
          \fig{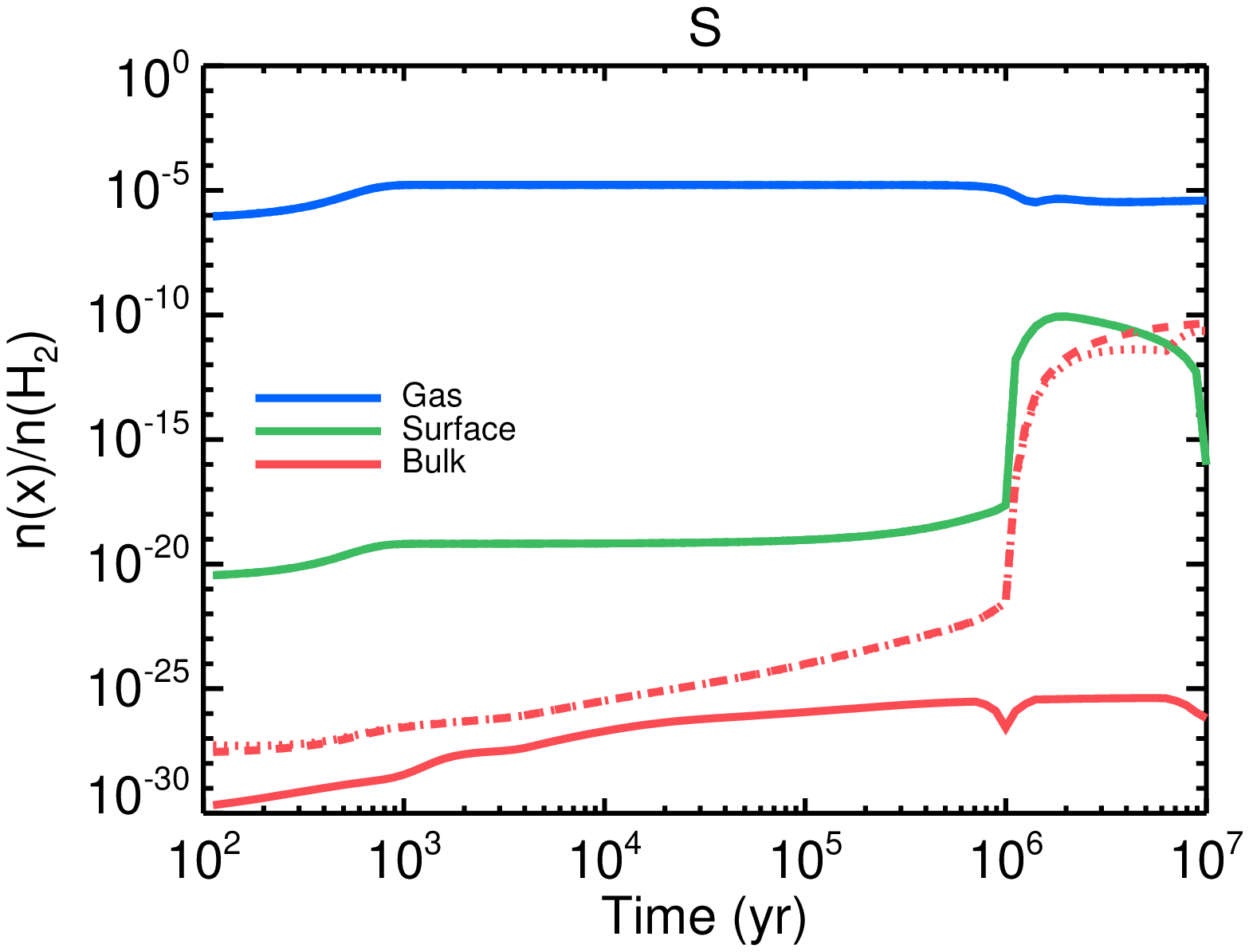}{0.5\textwidth}{}
          \fig{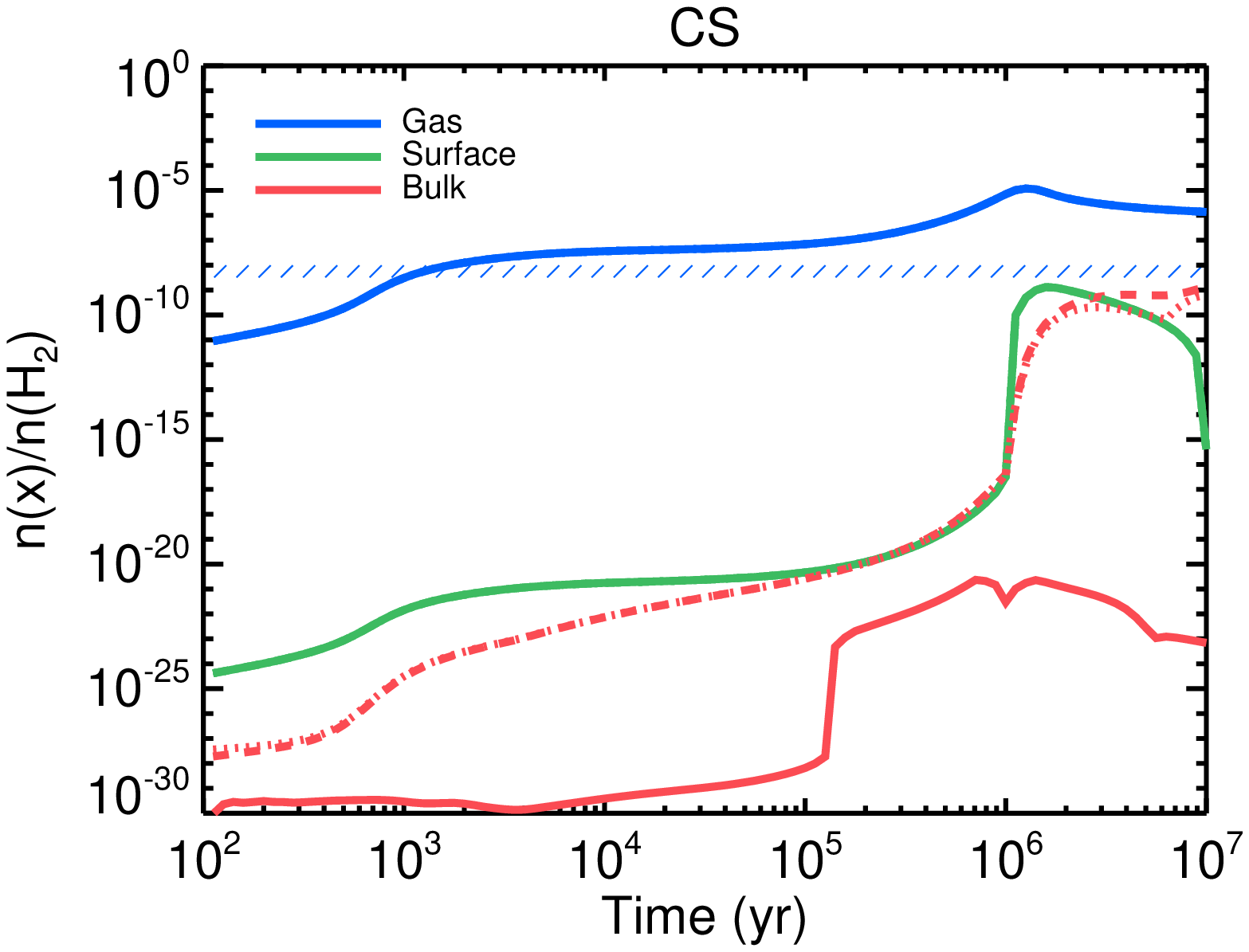}{0.5\textwidth}{}
}
\gridline{
          \fig{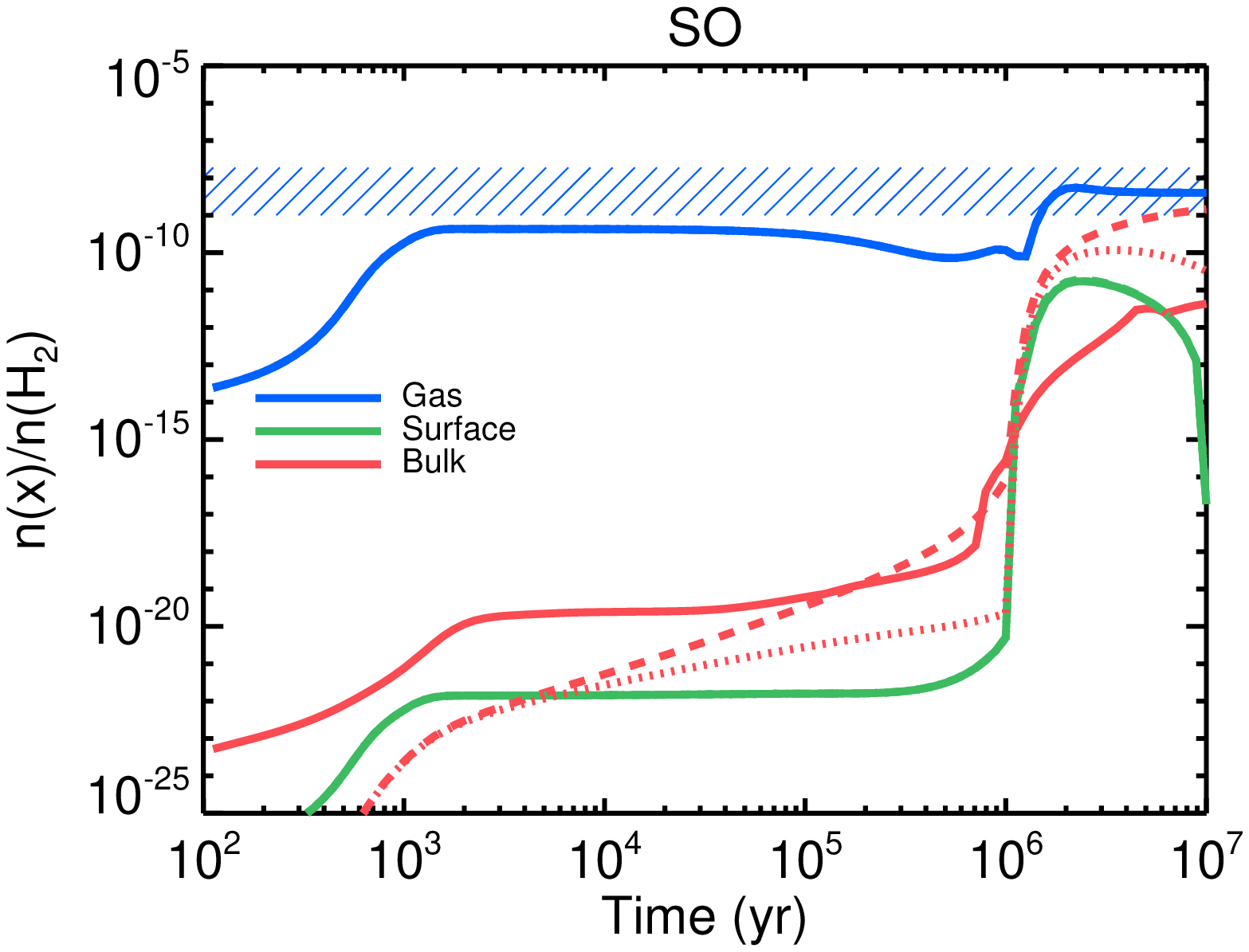}{0.5\textwidth}{}
          \fig{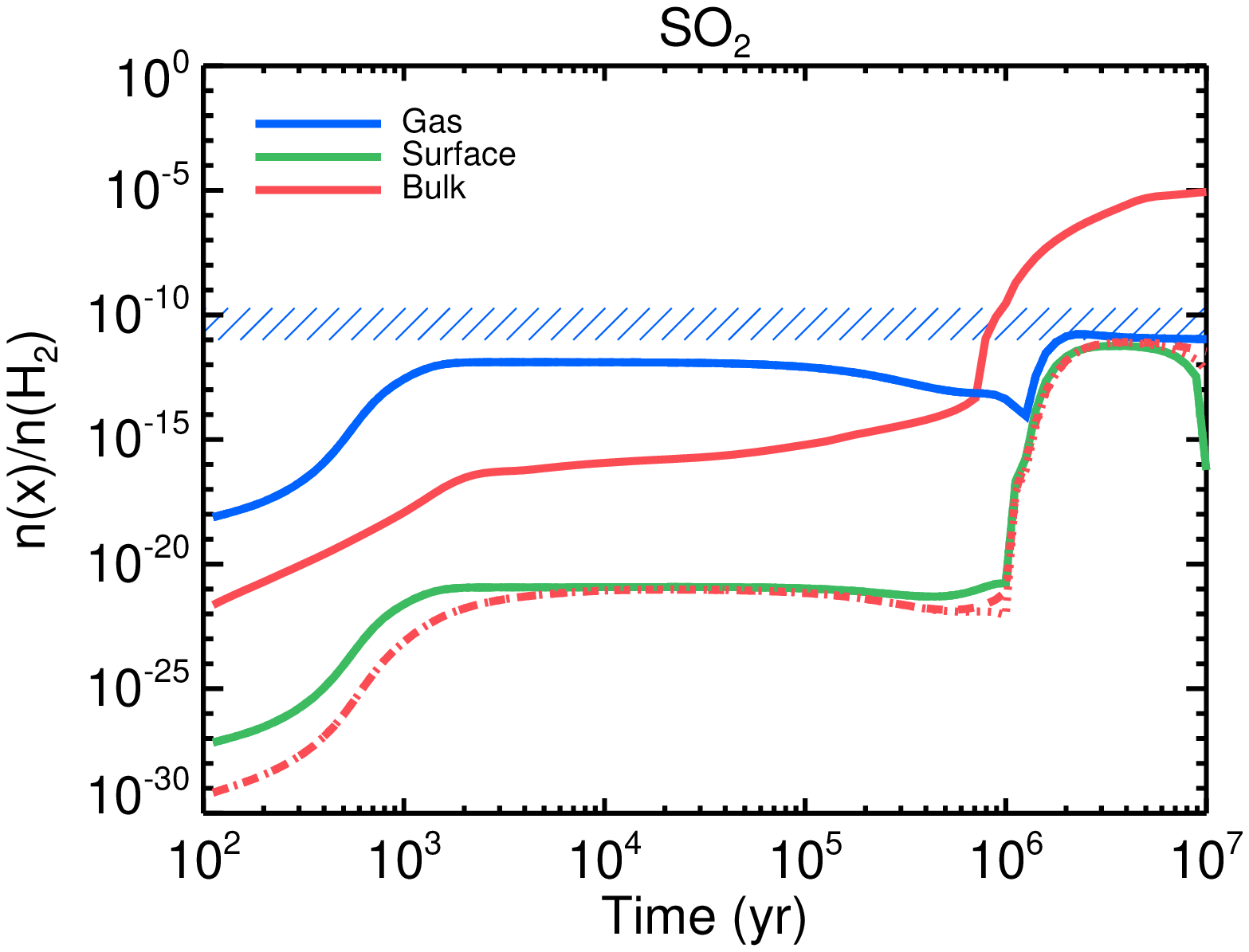}{0.5\textwidth}{}
}
\gridline{
          \fig{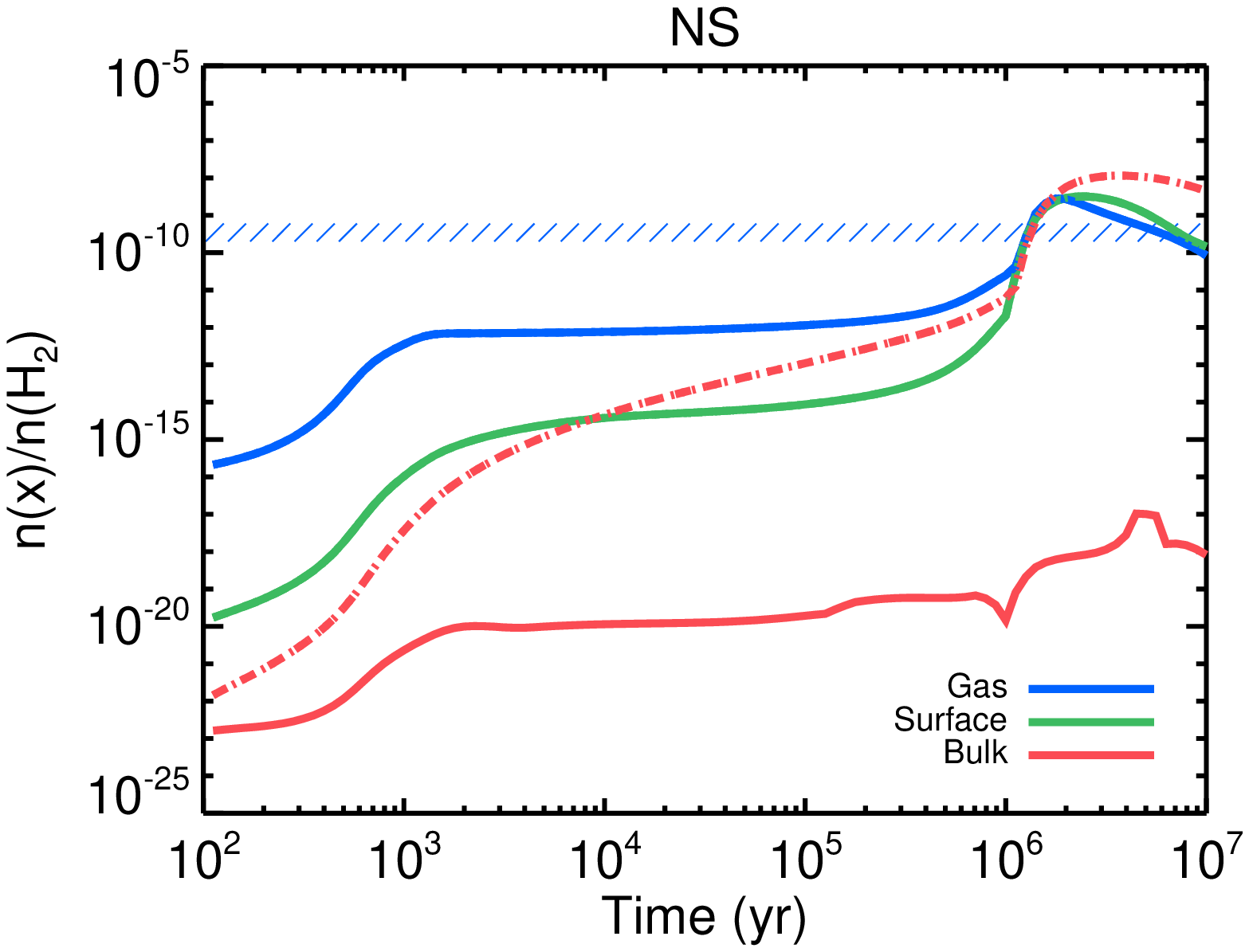}{0.5\textwidth}{}
          \fig{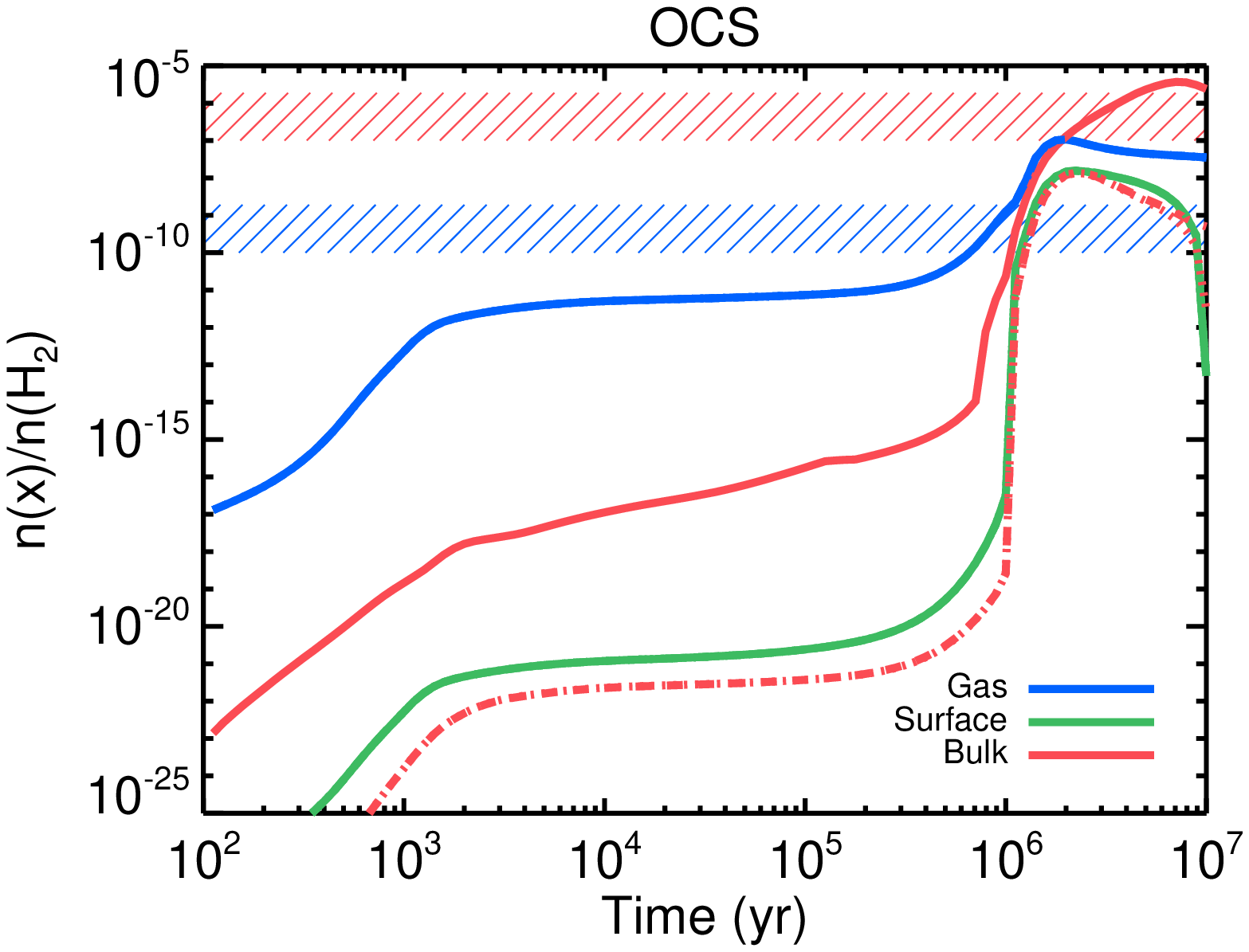}{0.5\textwidth}{}
}
\caption{ Calculated abundances of S, CS, \ce{SO}, \ce{SO2}, \ce{NS}, and \ce{OCS} in models C (solid line),  D (dashed line), and E (dotted line). Gas-phase observational abundances for dense clouds, where available, are represented by horizontal blue hatched bars (see \citet{laas_modeling_2019} and references therein.).
\label{fig:supblock2}}
\end{figure}

\begin{figure}
\gridline{
          \fig{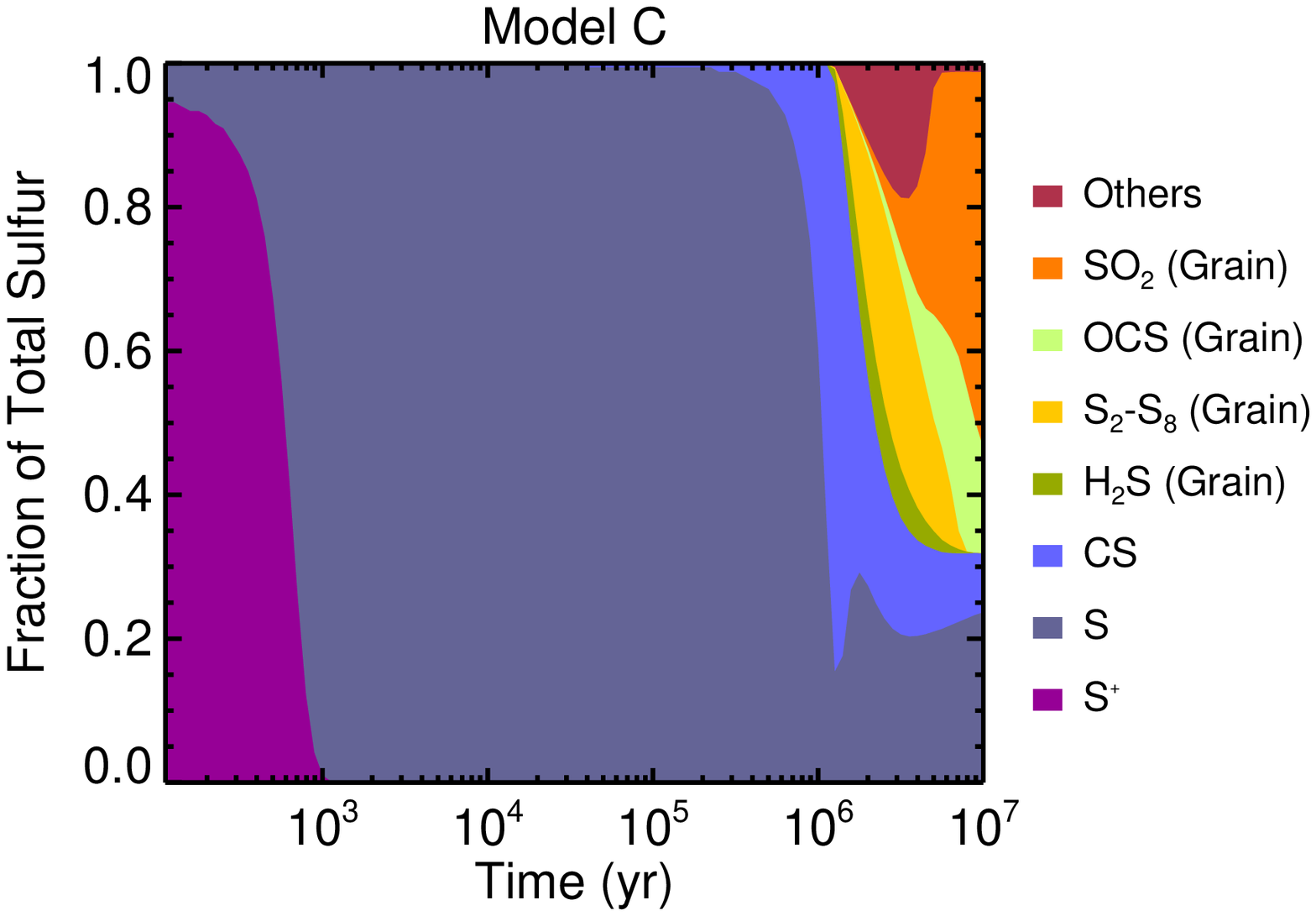}{0.5\textwidth}{}
          \fig{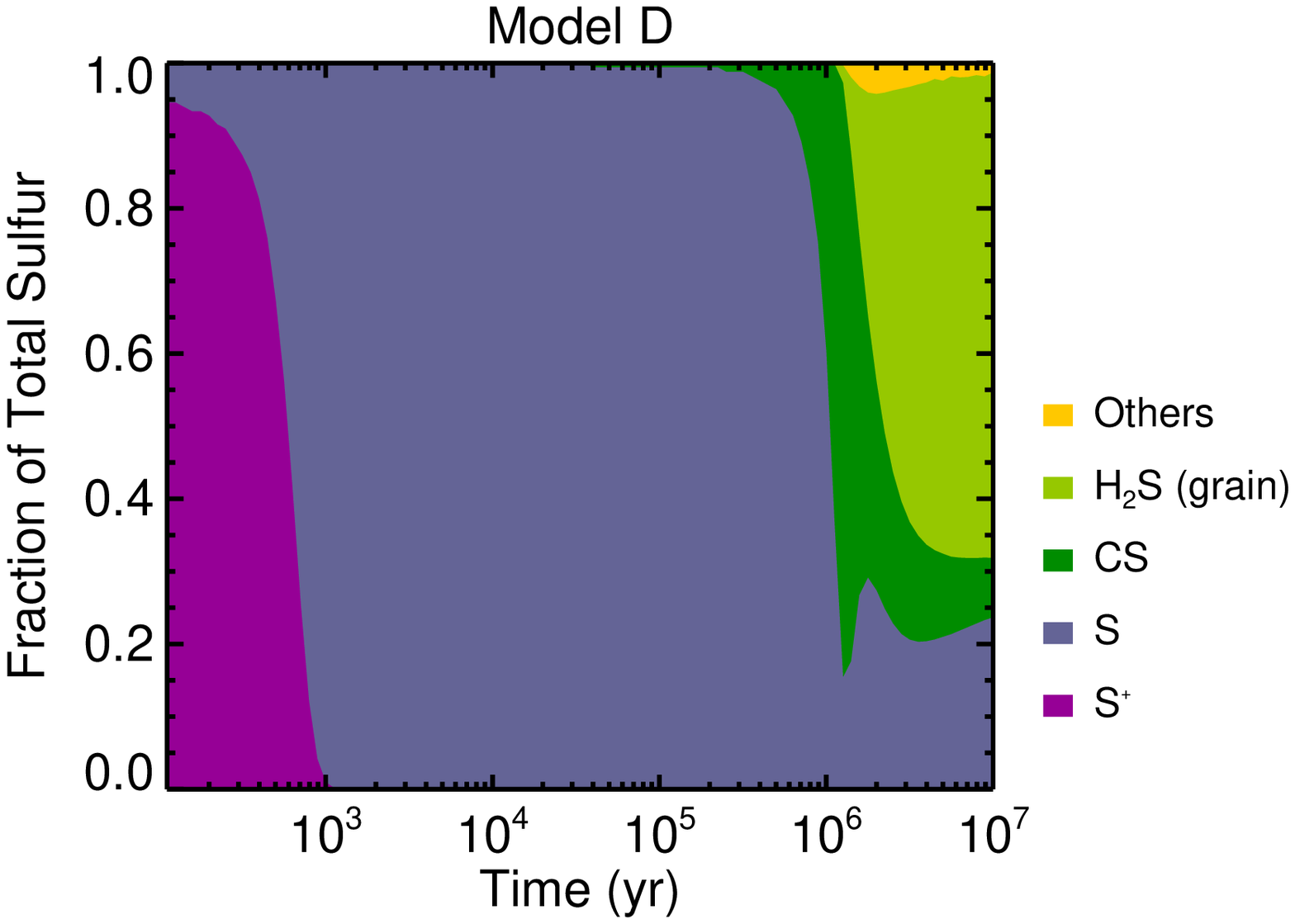}{0.5\textwidth}{}
}
\gridline{
          \fig{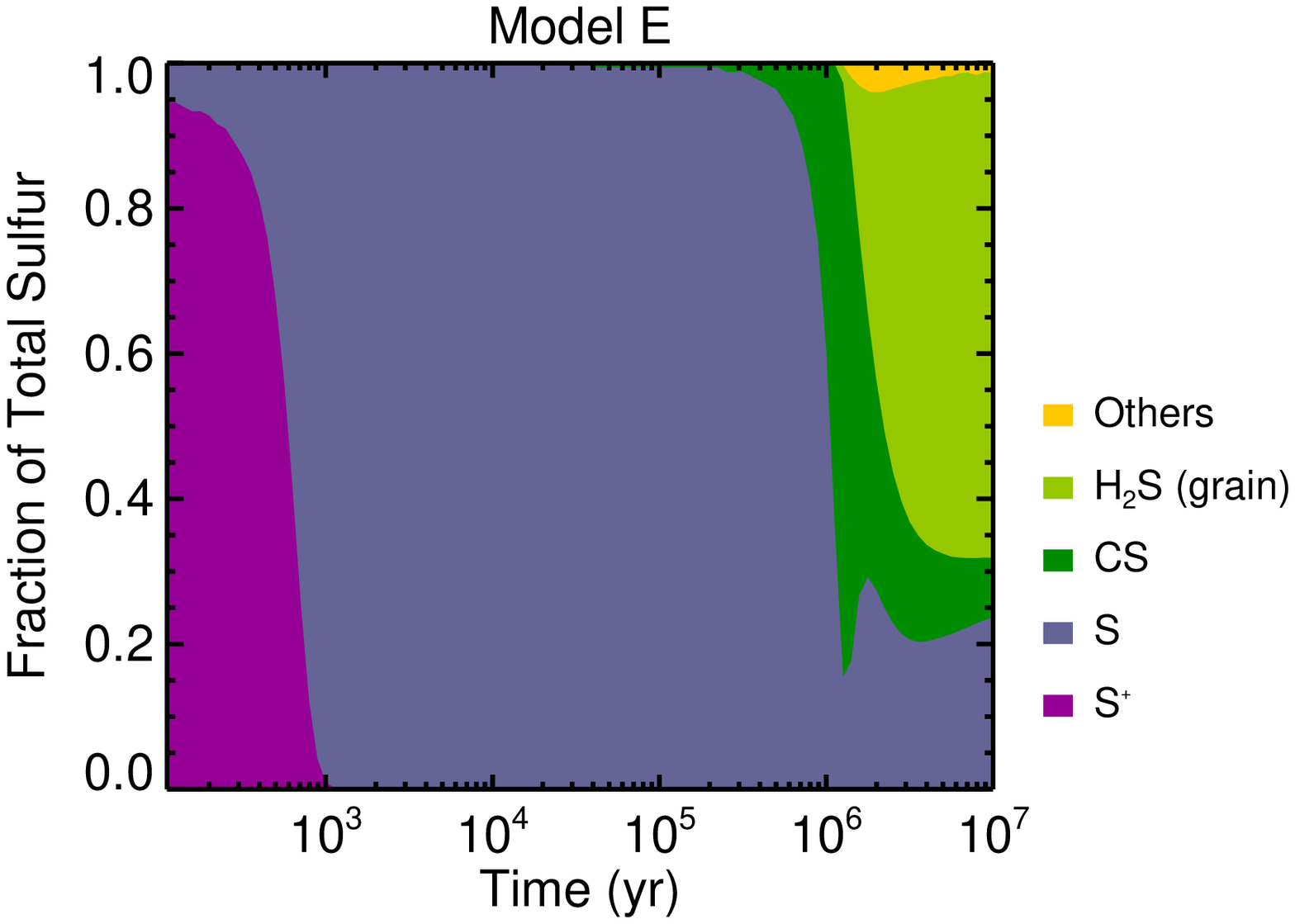}{0.5\textwidth}{}
}
\caption{Main sulfur-bearing species in Models C - E.
\label{fig:supblock4}}
\end{figure}

\FloatBarrier

\bibliography{bibliography}

\end{document}